\newif\ifmainpart

\mainpartfalse

\ifmainpart
\documentclass[sigplan,screen]{acmart} 

\copyrightyear{2022} 
\acmYear{2022} 
\setcopyright{rightsretained} 
\acmConference[PLDI '22]{Proceedings of the 43rd ACM SIGPLAN International Conference on Programming Language Design and Implementation}{June 13--17, 2022}{San Diego, CA, USA}
\acmBooktitle{Proceedings of the 43rd ACM SIGPLAN International Conference on Programming Language Design and Implementation (PLDI '22), June 13--17, 2022, San Diego, CA, USA}
\acmDOI{10.1145/3519939.3523714}
\acmISBN{978-1-4503-9265-5/22/06}

\else
\documentclass[acmsmall,nonacm,10pt]{acmart}
\fi

\ifmainpart
\newcounter{BalanceAtReference}
\newcounter{ReferenceIndexForBalancing}

\makeatletter

\global\@ACM@balancefalse

\def\@balancelastpageonce{%
  \ifnum\value{ReferenceIndexForBalancing}=\value{BalanceAtReference}
    \newpage
  \else
    \relax
  \fi
  \stepcounter{ReferenceIndexForBalancing}
}
\pretocmd{\bibitem}{\@balancelastpageonce}
  {} %
  {\@latex@error{Patching \bibitem failed}{\@ehd}}

\makeatother
\fi

\setcopyright{none}

\newcommand{\optional}[2]{%
\ifmainpart
#1\xspace
\else
#2\xspace
\fi
}

\usepackage{booktabs}   %

\usepackage{ifthen}

\usepackage{xspace}

\usepackage{tikz}
    \usetikzlibrary{
        arrows,
        arrows.meta,
        backgrounds,
        calc,
        chains,
        decorations,
        decorations.pathreplacing,
        decorations.pathmorphing,
        decorations.markings,
        fit,
        matrix,
        patterns,
        positioning,
        shadows,
        shapes}
\usepackage{pgfplots}
\usepackage{pgfplotstable}
\usepgfplotslibrary{units}

\usepackage{algorithm}
\usepackage[noend]{algpseudocode}

\usepackage{amsmath}

\usepackage{thmtools} 
\usepackage{thm-restate}

\usepackage{stmaryrd}

\usepackage{paralist}
\usepackage{microtype}

\usepackage{mathtools}

\usepackage{subcaption}
\usepackage{cleveref}

\usepackage{etoolbox}
\AtBeginEnvironment{algorithmic}{\small}

\usepackage{siunitx}

\declaretheorem[name=Property]{prop}	%

\definecolor{ao(english)}{rgb}{0.0, 0.5, 0.0}
\definecolor{royalblue(web)}{rgb}{0.25, 0.41, 0.88}

\newcommand{\setCommentColor}[1]{%
	\ifthenelse{\equal{#1}{bk}}%
		{\colorlet{colorVar}{red!50}}%
		{\ifthenelse{\equal{#1}{pv}}%
			{\colorlet{colorVar}{blue}}%
			{\ifthenelse{\equal{#1}{mg}}%
				{\colorlet{colorVar}{ao(english)}}%
			{\ifthenelse{\equal{#1}{jr}}%
				{\colorlet{colorVar}{magenta}}%
				{}%
			}%
		}%
	}%
}

\newcommand{\commentAuthor}[1]{%
	\ifthenelse{\equal{#1}{jr}}%
		{Jan:\ }%
		{\ifthenelse{\equal{#1}{cb}}%
			{Canberk:\ }%
			{\ifthenelse{\equal{#1}{xx}}%
				{thirdauthor:\ }%
			{\ifthenelse{\equal{#1}{zz}}%
				{fourthauthor:\ }%
				{}%
			}%
		}%
	}%
}

\newcommand{\blocks}{\textit{Block}}
\newcommand{\symblocks}{\textit{SymBlock}}
\newcommand{\lines}{\textit{Line}}
\newcommand{\setstate}{\textit{SetState}}
\newcommand{\symsetstate}{\textit{SymSetState}}
\newcommand{\cachestate}{\textit{CacheState}}
\newcommand{\symcachestate}{\textit{SymCacheState}}
\newcommand{\hierarchycachestate}{\textit{2LevelCacheState}}
\newcommand{\policystate}{\textit{PolicyState}}
\newcommand{\invalidline}{\epsilon}
\newcommand{\set}{\textit{Set}}

\newcommand{\cacheindex}{\textit{index}}
\newcommand{\updateset}{\textit{UpSet}}
\newcommand{\classifyset}{\textit{ClSet}}
\newcommand{\updatecache}{\textit{UpCache}}
\newcommand{\classifycache}{\textit{ClCache}}
\newcommand{\updatehierarchicalcache}{\textit{Up2LevelCache}}
\newcommand{\symupdatecache}{\textit{SymUpCache}}
\newcommand{\symupdateset}{\textit{SymUpSet}}
\newcommand{\symclassifycache}{\textit{SymClCache}}

\newcommand{\symupdatecacheit}{\textit{SymUpCache}}

\newcommand{\sem}[1]{\llbracket #1\rrbracket}

\newcommand{\conc}{\gamma}

\newsavebox{\verbbox} %

\begin{document}

\title{Warping Cache Simulation of Polyhedral Programs}
\optional{}{\titlenote{Extended version of PLDI 2022 paper.}}            %

\author{Canberk Morelli}
\orcid{0000-0002-5193-7915}             %
\affiliation{
  \institution{Saarland University}            %
  \streetaddress{Saarland Informatics Campus}
  \city{Saarbrücken} 
  \country{Germany}                    %
}
\email{s8camore@stud.uni-saarland.de}          %

\author{Jan Reineke}
\orcid{0000-0002-3459-2214}             %
\affiliation{
  \institution{Saarland University}            %
  \streetaddress{Saarland Informatics Campus}
  \city{Saarbrücken}
  \country{Germany}                   %
}
\email{reineke@cs.uni-saarland.de}         %

\begin{abstract}
Techniques to evaluate a program's cache performance fall into two camps:
1.~Traditional trace-based cache simulators precisely account for sophisticated real-world cache models and support arbitrary workloads, but their runtime is proportional to the number of memory accesses performed by the program under analysis. 
2.~Relying on implicit workload characterizations such as the polyhedral model, analytical approaches often achieve problem-size-independent runtimes, but so far have been limited to idealized cache models.
 
We introduce a hybrid approach, warping cache simulation, that aims to achieve applicability to real-world cache models and problem-size-independent runtimes. As prior analytical approaches, we focus on programs in the polyhedral model, which allows to reason about the sequence of memory accesses analytically. Combining this analytical reasoning with information about the cache behavior obtained from explicit cache simulation allows us to soundly fast-forward the simulation. By this process of {warping}, we accelerate the simulation so that its cost is often independent of the number of memory accesses.
\end{abstract}

\ccsdesc[500]{Software and its engineering~Software performance}
\ccsdesc[500]{Software and its engineering~Automated static analysis}
\keywords{cache model, simulation, performance analysis, data independence}  %

\maketitle

\section{Introduction}

Traditionally, the efficiency of an algorithm has been determined by evaluating its time complexity. 
Today, evaluating an algorithm's cache performance has become equally important.
Over the past thirty years, the increasing processor-memory gap has led to the introduction of complex memory hierarchies consisting, in particular, of multiple cache levels.
As a consequence, a program's runtime on modern hardware heavily depends on how well it exploits the underlying memory hierarchy. 
However, unlike time complexity, cache performance cannot easily be gauged in a compositional manner from a program's parts, i.e., the composition of two cache-efficient parts may be cache inefficient, and vice versa.

This calls for automatic methods to evaluate a program's cache performance, to inform programmers and compilers so that they can make informed choices about data-locality transformations. 
Cache performance analysis has already received considerable attention.
Prior work can roughly be divided into two camps:\looseness=-1

1. \emph{Traditional cache simulators}, such as Dinero~IV~\cite{Edler1999} or CASPER~\cite{Iyer2003}, simulate a program's cache behavior by explicitly iterating over the trace of memory accesses generated by the program. 
		The advantage of this approach is that it is applicable to arbitrary workloads and it is possible to precisely model modern memory hierarchies, including sophisticated cache replacement policies, such as Pseudo-LRU~\cite{al04} or Quad-age LRU~\cite{jaleel10,jahagirdar12} found in real-world microarchitectures~\cite{Vila2020,Abel2020}.
		The main drawback of traditional simulators is that their runtime is \emph{proportional to the number of memory accesses} a program performs. As a consequence, the simulation of programs operating on large amounts of data may take weeks or more.
		
2. \emph{Analytical cache models}~\cite{Ghosh1997,Ghosh1999,Chatterjee2001,Vera2002,Vera2004,Cascaval2003,Beyls2005,Bao2018,Gysi2019}, on the other hand, e.g. PolyCache~\cite{Bao2018} or HayStack~\cite{Gysi2019}, aim to achieve analysis times that are \emph{independent of the number of memory accesses} performed by the program under analysis.
		To this end, they rely on implicit representations of a program's memory accesses. 
		A prominent such program representation is the \emph{polyhedral model}~\cite{Feautrier1991,Benabderrahmane2010}, which, loosely speaking, captures a program's memory accesses as polyhedra.
		For such programs, the number of cache misses can be obtained analytically by applying a sequence of algebraic operations on the program representation and by applying symbolic counting techniques.  
		One main drawback of these analytical models is that they are limited to simplified cache models: HayStack~\cite{Gysi2019} applies to inclusive hierarchies of fully-associative caches with least-recently-used (LRU) replacement; PolyCache~\cite{Bao2018} applies to hierarchies of set-associative caches but is also limited to LRU replacement and can handle non-write allocate caches only approximately. %

In this paper, we introduce a new approach called \emph{warping cache simulation} that aims to combine the strengths of traditional cache simulators and analytical cache models.
Warping cache simulation is applicable to realistic models of modern memory hierarchies, supporting hierarchical caches with various write policies and arbitrary replacement policies, and its runtime is often independent of the number of memory accesses of a program.

At its core, warping exploits the following data-indepen-dence property of caches:
Assume $c_1$ and $c_2$ are two cache states that are equal up to a renaming of the addresses of the cached memory blocks, i.e., there is a bijection $\pi$ mapping the memory blocks of $c_1$ to those of $c_2$, such that $\pi(c_1) = c_2$.
Then, an access to~$c_1$ under block $b$ is a cache hit if and only if $\pi(b)$ hits in state $c_2$.
Similarly, the resulting cache states $c_1', c_2'$ under accesses to $b$ and $\pi(b)$ are guaranteed to be related to each other under the same bijection $\pi$.

\begin{figure}[t]
\begin{center}
\begin{verbatim}
    for (int i = 1; i < 999; i++)
        B[i-1] = A[i-1] + A[i];
\end{verbatim}
\vspace{2mm}
\includegraphics[width=7.5cm]{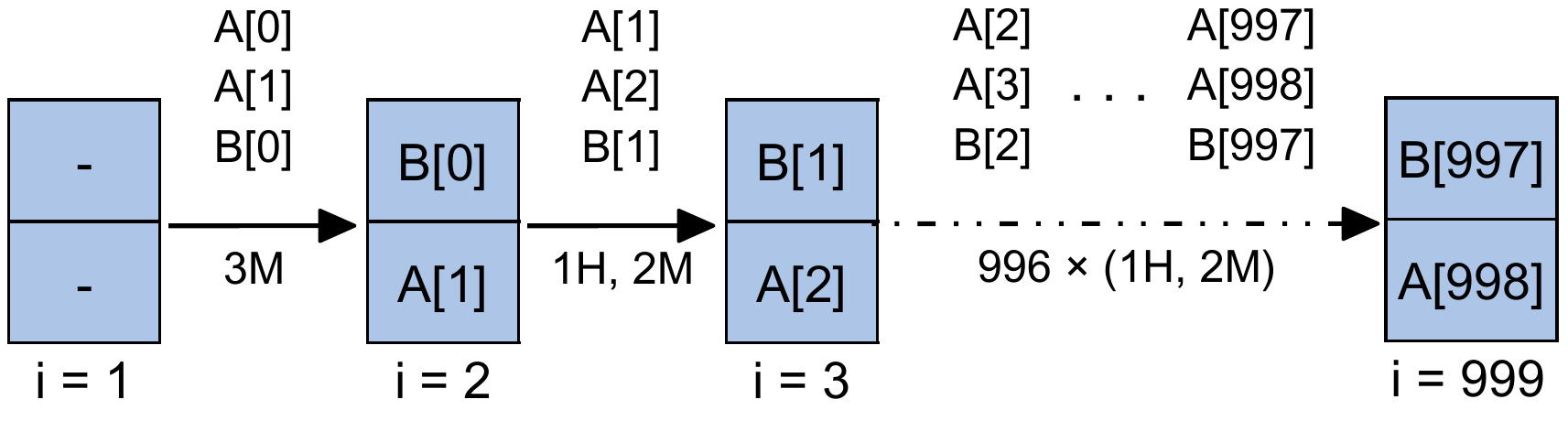}
\end{center}
\caption{1D stencil computation and its warping simulation.\label{fig:runningexample}}
\end{figure}

\begin{figure}[t]
\begin{center}
\includegraphics[width=7.5cm]{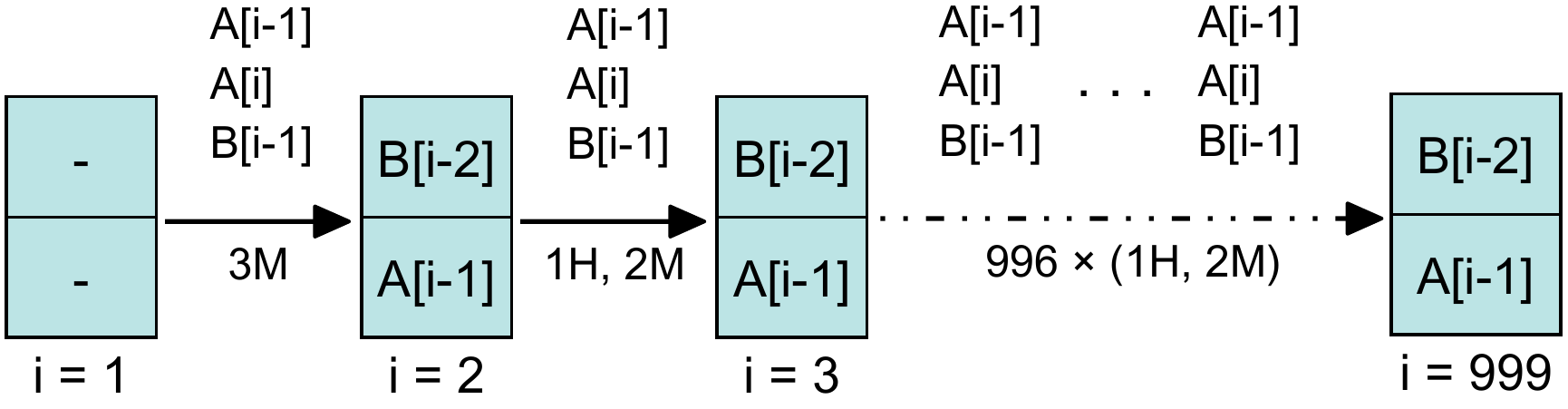}
\end{center}
\caption{Symbolic warping cache simulation.\label{fig:runningexamplesymbolic}}
\end{figure}

Let us illustrate how we can exploit this data-independence property in warping cache simulation at the hand of the 1D stencil computation in Figure~\ref{fig:runningexample}, which will serve as a running example throughout the paper.
In the example, we assume a small fully-associative cache of size two with least-recently-used (LRU) replacement and that each array cell occupies a full cache line; but the approach equally applies to more complex real-world caches.
In each loop iteration, the program accesses $A[i], A[i-1]$ and $B[i-1]$.
Thus, after the first loop iteration, which results in three misses, $B[0]$ and $A[1]$ are cached.
All subsequent iterations will hit on the access to $A[i-1]$ because it was cached in the previous iteration.
Iteration~$i$ results in a cache state containing $B[i-1]$ and $A[i]$.
Thus cache states in consecutive iterations are related under the simple bijection that maps memory block~$i$ to memory block~$i+1$.
Warping cache simulation detects this relation. %
Then, it checks whether the future memory accesses relate to the past memory accesses in the same way that the matching cache states relate to each other.
If that is the case, the simulation may fast forward, potentially all the way to the end of the loop, determining the resulting number of cache misses and the final cache state analytically.
In our example, warping simulation fast forwards through the entire loop after explicitly simulating the loop for two iterations.\looseness=-1

To make this basic idea a reality, we introduce {symbolic cache simulation} to efficiently determine whether two cache states encountered during concrete simulation are related under a bijection.
Figure~\ref{fig:runningexamplesymbolic} illustrates the symbolic cache simulation of the 1D stencil computation.
In our example, symbolic cache simulation determines that the cache states obtained after the first and the second iteration both contain $A[i]$ and $B[i-1]$ for different values of the loop iterator~$i$.
Hashing the symbolic cache states obtained in different iterations allows to efficiently detect such a match, also across several iterations.
Further, we employ polyhedral techniques to check whether future memory accesses satisfy the warping conditions implied by matching cache states obtained during simulation.
We have implemented our approach and applied it to the PolyBench~\cite{Pouchet2012} benchmark suite.
Our experiments show that warping cache simulation may outperform traditional cache simulation by several orders of magnitude.

To summarize, we make the following contributions:
\begin{itemize}
	\item We introduce \emph{warping cache simulation}, the first approach that is both applicable to real-world cache architectures and may achieve simulation times that are independent of the number of memory accesses performed by the program.\looseness=-1
	\item We implement warping cache simulation and experimentally evaluate its performance, demonstrating that warping cache simulation may outperform traditional cache simulation by several orders of magnitude.
\end{itemize}

\section{Caches and Data Independence}
    Caches are fast but small memories that buffer parts of the large but slow main memory in order to bridge the speed gap between the processor and main memory.
    Caches operate at the granularity of memory blocks $b \in \blocks$, which are stored in the cache in \emph{cache lines} of the same size.
    In order to facilitate an efficient cache lookup, the cache is organized into \emph{sets} such that each memory block maps to a unique cache set.
    The size~$k$ of a cache set is called the \emph{associativity} of the cache.
    If an accessed block resides in the cache, the access \emph{hits} the cache.
    Upon a cache \emph{miss}, the block is loaded from the next level of the memory hierarchy, e.g. main memory in case of the last-level cache. 
    If the corresponding cache set is full, another memory block is evicted to make place for the newly loaded block.
    The block to evict is determined by the \emph{replacement policy}.

    To ease the formal development, we first formalize the behavior of individual cache sets, which is already sufficient to capture fully-associative caches; and then generalize this formalization to set-associative caches.

\subsection{Cache Sets}

The state of an individual cache set in a cache of associativity~$k$ is a pair 
\begin{equation*}
	s \in \setstate = (\lines \rightarrow (\blocks \cup \{\invalidline\})) \times \policystate,
\end{equation*}
where $\lines = \{1, \dots, k\}$, and thus the first component of the pair captures the memory blocks stored in the $k$ cache lines of the cache set.
Empty lines are represented by $\invalidline$.
The state of many replacement policies can be fully encoded in the order in which memory blocks occupy a set's cache lines. 
Examples of such policies are least-recently-used (LRU), first-in first-out (FIFO), and Pseudo-LRU (PLRU)~\cite{Abel2013}.
For such policies, a separate $\policystate$ component then becomes unnecessary and can be omitted. 
E.g. under LRU, cache lines can be ordered from most- to least-recently-used, adapting the order upon each access.
Similarly, under FIFO cache lines can be ordered from last- to first-in.
Such encodings are slightly more complicated for PLRU and discussed e.g. in \cite{Grund2010b,Doychev15}. 
To model more complex policies, such as Quad-age LRU~\cite{jaleel10,jahagirdar12}, the $\policystate$ component is used to capture additional state of the replacement policy. 
Given a set state $s = (m,ps)$, we refer to the mapping of $s$ by $s.m$ and to the policy state of $s$ by $s.ps$.\looseness=-1

Note that this model does not include the data stored in the cache set as this is not relevant to determine whether a memory access results in a hit or a miss. 
For simplicity, we also do not differentiate between reads and writes, which may make a difference depending on the write policy.
Thus the formalization applies to write-allocate caches, but our implementation also supports no write-allocate caches.

We may model the effect of a memory access on the cache state using the two functions $\updateset: \setstate \times \blocks \rightarrow \setstate$ and $\classifyset : \setstate \times \blocks \rightarrow \mathbb{B}$, 
which take as input a set state and the accessed memory block and return the updated set state and the access's classification as a hit or a miss, respectively. 

The definition of $\updateset$ depends on the particular replacement policy.
For LRU, e.g., it is defined as follows:
\begin{equation*}
	\updateset_{LRU}(s, b) := \lambda l \in \lines. \begin{cases}
									b		& : \textit{if } l =1\\
									s(l)		& : \textit{if } \exists l' < l: s(l') = b\\
									s(l-1)		& : \textit{otherwise}
								\end{cases}
\end{equation*}

Our approach is applicable to \emph{any} replacement policy as long as it satisfies the data-independence property we will define shortly.
In contrast, $\classifyset$ can be defined generically by inspecting the contents of the cache lines:
\begin{equation}
	\classifyset(s, b) := \begin{cases}
					\textit{true}	& : \textit{if } \exists i: s.m(i) = b\\
					\textit{false}	& : \textit{otherwise}
				\end{cases}
\end{equation}

Let $\Pi \subset \blocks \rightarrow \blocks$ be the set of bijections from memory blocks to memory blocks.
A bijection~$\pi \in \Pi$ can be applied to a set state as follows:
 \[\pi(s) := (\lambda l. \pi(s.m(l)), s.ps),\]
where we define $\pi(\epsilon) = \epsilon$.
In other words, the bijection is applied to the contents of each cache line, mapping empty lines to empty lines.

\begin{prop}[Data independence of cache sets]\label{prop:dataindependence}
Let $s \in \setstate$, $b \in \blocks$, and $\pi \in \Pi$. Then:
\begin{align}\label{eq:dataindependenceset}
	\pi(\updateset(s,b)) & = \updateset(\pi(s), \pi(b))
\end{align}	
\end{prop}
In other words, the cache update %
is independent of the particular memory blocks stored in a cache set.
To simplify the following statements, we do not restate Property~\ref{prop:dataindependence} in the remainder of the paper, but implicitly assume it holds.

All cache architectures we are aware of satisfy Property~\ref{prop:dataindependence}.
Recent measurement-based approaches~\cite{Abel2013,Abel2020,Vila2020} to automatically derive cache models are also naturally limited to models satisfying data independence.
Our warping cache simulator supports LRU, FIFO, PLRU~\cite{al04}, and Quad-age LRU~\cite{jaleel10,jahagirdar12}, which allows to model the L1 and L2 caches of most recent Intel microarchitectures~\cite{Abel2020,Vila2020}.
Other policies can be added as long as they satisfy data independence.\looseness=-1

\subsection{Set-associative Caches}

Set-associative caches can be seen as the composition of multiple cache sets.
Typically the number of cache sets $s$ is a power of two, so that the cache set that a memory block maps to is determined by a subset of its address, which is commonly referred to as the cache index.

In the following, we model the mapping from memory blocks into cache sets using the function $\cacheindex: \blocks \rightarrow \set$, where $\set = \{0, \dots, s-1\}$.
Most real-world caches employ a modulo mapping of blocks to cache sets, i.e., $\cacheindex(b) = b \bmod s$.
Then, the state of a set-associative cache can be captured simply as a mapping from cache sets to their states:
$c \in \cachestate = \set \rightarrow \setstate$.

A memory access results in a hit if it hits in the cache set that it maps to:
\begin{equation}\label{eq:defclassifycache}
\classifycache(c,b) := \classifyset(c(\cacheindex(b)), b)
\end{equation} 

Cache states are updated by updating the cache set the block maps to:
\begin{align}\label{eq:defupdatecache}
	\updatecache(c, b) & := c[\cacheindex(b) \mapsto \updateset(c(\cacheindex(b)), b)]
\end{align}
\newcommand{\Piindex}{\Pi_{\cacheindex_=}}
\newcommand{\PiindexOne}{\Pi_{\cacheindex_{=,1}}}
\newcommand{\PiindexTwo}{\Pi_{\cacheindex_{=,2}}}
Thus, cache sets are updated independently of each other, which implies another source of symmetry we seek to exploit.
To this end, let $\Piindex$ be the set of bijections on blocks that preserve the partition of blocks into cache sets:\looseness=-1
\begin{multline*}
	\Piindex := \{\pi \in \Pi \mid \forall b, b' \in \blocks: (\cacheindex(b) = \cacheindex(b'))\optional{\\}{} \Leftrightarrow (\cacheindex(\pi(b)) = \cacheindex(\pi(b')))\}
\end{multline*}
A bijection $\pi \in \Piindex$ induces a bijection $\pi_\set$ on cache sets:
\begin{equation*}
\pi_\set := \{(\cacheindex(b), \cacheindex(\pi(b))) \mid b \in \blocks\}.
\end{equation*}
This allows to apply bijections from $\Piindex$ to cache states:
\begin{equation}\label{eq:defpicachestate}
\pi(c) := \lambda s. \pi(c(\pi_\set^{-1}(s)))
\end{equation}

Assuming Property~\ref{prop:dataindependence} holds on the underlying cache sets, it also holds for the resulting set-associative cache:
\begin{restatable}[Data independence of caches]{thm}{dataindependence}\label{thm:dataindependence}
Let $c \in \cachestate$, $b \in \blocks$, and $\pi \in \Piindex$. Then:
\begin{align}
	\pi(\updatecache(c,b)) &= \updatecache(\pi(c), \pi(b)),\label{eq:dataindependence}\\
	\classifycache(c,b) &= \classifycache(\pi(c), \pi(b)).\label{eq:dataindependenceclassification}
\end{align}
\end{restatable}
The proof of this theorem and all other proofs are given in the \optional{extended version of this paper~\cite{arxivVersion}}{appendix}.

\begin{figure}[t]
\begin{center}
\optional{\includegraphics[width=\linewidth]{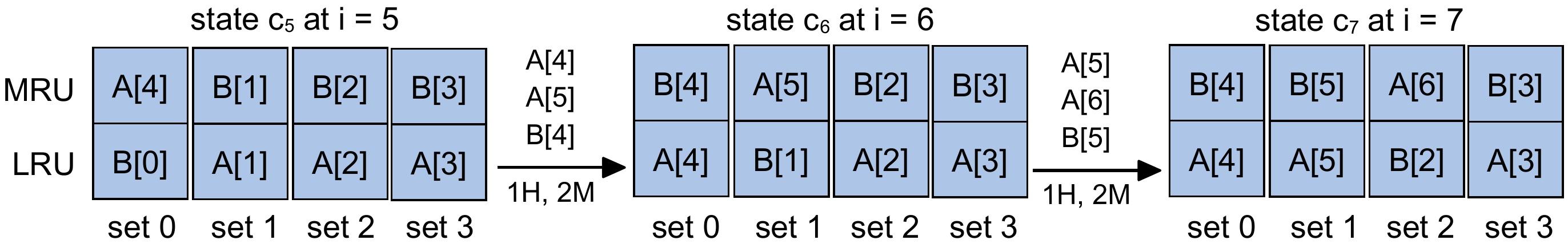}}{\includegraphics[width=0.7\linewidth]{figures/ex-set-conc}}
\end{center}
\caption{Example illustrating the data independence of set-associative caches.\label{fig:runningexamplesetassociative}}
\end{figure}

\begin{example}
Let us illustrate Theorem~\ref{thm:dataindependence} at the hand of the 1D stencil code from Figure~\ref{fig:runningexample}.
Assume a set-associative cache consisting of four cache sets of associativity two with LRU replacement.
Further assume, as in the previous examples, that each array cell occupies one full cache line and that the index of both $A[0]$ and $B[0]$ is zero. 
Then, the execution reaches cache state~$c_5$ in Figure~\ref{fig:runningexamplesetassociative} at the start of loop iteration~$5$.
In the figure, the cache lines within each cache set are ordered from most-recently-used (MRU) to least-recently-used (LRU).
Performing the accesses of iteration~5 yields cache state~$c_6$, with $c_6 = \pi(c_5)$, where $\pi(i) = i+1$, and thus $\pi_\set(s) = (s + 1) \bmod 4$.
As the accesses in iteration~6 relate to those of the iteration~5 under the same bijection~$\pi$, following Theorem~\ref{thm:dataindependence}, the next state can be obtained as $c_7 = \pi(c_6)$.

\end{example}

\subsection{Multi-level Caches}

Memory hierarchies of modern multi-core processors contain multiple cache levels.
Typically, L1 and L2 caches are private to a core, while the L3 cache is shared among all of the processor's cores.

Cache hierarchies are governed by \emph{inclusion policies}~\cite{Solihin2015} that determine how the contents of a given cache level relate to those of the next level of the hierarchy.
In the \optional{extended version~\cite{arxivVersion}}{appendix}, we show how to model a two-level \emph{non-inclusive non-exclusive} cache hierarchy and prove that this model satisfies data independence.
We note that \emph{inclusive} and \emph{exclusive} cache hierarchies also satisfy data independence and could be captured in a similar manner.

\section{Polyhedral Program Representation}

The polyhedral model~\cite{Feautrier1991,Feautrier1992,Benabderrahmane2010}
is a mathematical framework to succinctly describe and manipulate programs' control flow and data-access patterns using Presburger arithmetic~\cite{Haase2018}.
In this section, we introduce a simple program representation resembling abstract syntax trees that is tailored to cache simulation.

\newcommand{\vi}{\vec{i}}
\newcommand{\vj}{\vec{j}}
\newcommand{\vk}{\vec{k}}
\newcommand{\vf}{\vec{f}}
\newcommand{\lexmin}{\textit{lexmin}\xspace}
\newcommand{\lexmax}{\textit{lexmax}\xspace}

\newcommand{\isl}{\textit{isl}\xspace}

\subsection{Presburger Sets and Maps}

To manipulate integer sets and to represent programs in the polyhedral model, we make use of \isl, the \emph{integer set library}~\cite{Verdoolaege2010}.
Here, we introduce how integer sets can be defined and some important operations on integer sets provided by \isl.
Our presentation loosely follows the tutorial by Verdoolaege~\cite{Verdoolaege2016}.
More details can be found there.

A \emph{Presburger set} $S = \{(i_1, \dots, i_n) \mid c\}$ is an integer set, i.e., a set of integer tuples $(i_1, \dots, i_n) = \vi \in \mathbb{Z}^n$, whose elements satisfy the Presburger formula $c$. 
The only free variables allowed in $c$ are $i_1, \dots, i_n$, so that the set~$S$ corresponds to the satisfying assignments of $c$.

Presburger formulas are first-order formulas that are limited to the Presburger language, which allows for addition~$+$, subtraction~$-$, integer constants~$d$, floored division by integer constants~$\lfloor \cdot/d\rfloor$ and the binary predicate~$\leq$.

\begin{example}
The set
	$E = \{(i, j) \mid \exists k: \lfloor i/7\rfloor=k+k \wedge i \leq j \}$
consists of all pairs of integers $(i, j)$, s.t. the floored division of $i$ by seven is even and for which $i \leq j$.
\end{example}

To ease notation, several other operations are supported as syntactic sugar, in particular multiplication by constants, modulo with a constant divisor, and comparison of integer tuples by lexicographic ordering~$\preceq$.
It is also convenient to refer to previously defined Presburger sets within the definition of a new set:
$F = \{(i,j) \mid \exists k: (i,k) \in E \wedge j+j = k\}$.

A \emph{Presburger relation} $R = \{(i_1, \dots, i_n) \rightarrow (j_1, \dots, j_m) \mid c\}$
relates integer tuples $\vi \in \mathbb{Z}^n$ to integer tuples $\vj \in \mathbb{Z}^m$, where the constraint $c$ has the same restrictions as in the case of Presburger sets.
For a Presburger relation $R$, $R_\textit{dom}$ denotes its domain, i.e., $R_\textit{dom} = \{\vi \mid \exists \vj: \vi \rightarrow \vj \in R\}$. 
Dually, 
$R_\textit{ran}$ denotes its range, i.e., $R_\textit{ran} = \{\vj \mid \exists \vi: \vi \rightarrow \vj \in R\}$. 
We express the lexicographic minimum and maximum of sets as $\lexmin(S) = \{\vi \mid \vi \in S \wedge \not\exists \vj \in S: \vj \prec \vi\}$ and $\lexmax(S) = \{\vi \mid \vi \in S \wedge \not\exists \vj \in S: \vi \prec \vj\}$, respectively. 

Given the definition of a set or relation we may use \isl to check whether the set is empty, and, if not, to extract an element of this set or relation.

\subsection{Static Control Parts}\label{sec:scops}

Warping cache simulation applies to \emph{static control parts} (SCoPs) of programs. 
SCoPs are loop nests whose control flow and memory-access behavior is determined statically, and thus independent of the program's inputs. Further restrictions apply to the loop bounds and the array index expressions, which are limited to affine expressions.
We also refer to such loops as \emph{polyhedral programs}.

For the purpose of cache simulation, it is sufficient to capture a SCoP's memory-access behavior.
Thus, we can safely abstract from the computations performed by a SCoP.
To this end, we introduce a tree-structured representation for SCoPs, which resembles the program's abstract syntax tree. %
This tree representation consists of two types of nodes:\looseness=-1
\begin{compactenum}	
	\item \emph{Loop nodes} correspond to loops in the source program. 
	\item \emph{Access nodes} form the leaves of the tree and correspond to the array accesses\footnote{If necessary, scalar variables can be modeled as zero-dimensional arrays.} performed by the program.
\end{compactenum}

\newcommand{\domain}{\textit{dom}}
\newcommand{\children}{\textit{children}}
\newcommand{\access}{\textit{access}}

\begin{lrbox}{\verbbox}
\begin{minipage}{.6\linewidth}\small
\begin{verbatim}
for (int i = 0; i < 100; i++) {
  c[i] = 0;
  for (int j = i; j < 100; j++) {
    c[i] = c[i] + A[i][j] * x[j];
  }
}
\end{verbatim}
\end{minipage}
\end{lrbox}%
\tikzset{%
    n/.style={draw, circle, thick, inner sep=1mm, minimum width=7mm},
    l/.style={draw, rounded corners=1mm, thick, inner sep=1mm},
    ll/.style={draw, rounded corners=1mm, shape=rectangle split, rectangle split parts=2, thick, inner sep=1mm, font=\vphantom{Q}}, %
    e/.style={shorten >=1mm, shorten <=1mm, thick, ->, >=stealth},
    i/.style={initial, initial text={}},
    every initial by arrow/.style={thick, ->, >=stealth}
} 

\begin{figure}[t]
\begin{tabular}{lr}
\usebox{\verbbox} & \parbox[c]{6.5cm}{~\hspace{-6mm}\begin{tikzpicture}[node distance=14mm, transform shape,every node/.style={scale=0.85}]
            	\node[draw, thick, rectangle, rounded corners, inner sep=2pt,minimum size=5mm, fill=white] (Li) {$L_i$};
            	\node[draw, thick, rectangle, rounded corners, inner sep=2pt, minimum size=5mm, fill=white, below of=Li, yshift=4mm,xshift=-6mm] (Aic) {$A_i$};
            	\node[draw, thick, rectangle, rounded corners, inner sep=2pt, minimum size=5mm, fill=white, below of=Li, yshift=4mm,xshift=6mm] (Lj) {$L_j$};
            	\node[draw, thick, rectangle, rounded corners, inner sep=2pt, minimum size=5mm, fill=white, below of=Lj, xshift=-2.0cm, yshift=4mm] (Aijc) {$A_{ij,1}$};
            	\node[draw, thick, rectangle, rounded corners, inner sep=2pt, minimum size=5mm, fill=white, right of=Aijc,xshift=-3mm] (AijA) {$A_{ij,2}$};
            	\node[draw, thick, rectangle, rounded corners, inner sep=2pt, minimum size=5mm, fill=white, right of=AijA,xshift=-3mm] (Aijx) {$A_{ij,3}$};
            	\node[draw, thick, rectangle, rounded corners, inner sep=2pt, minimum size=5mm, fill=white, right of=Aijx,xshift=-3mm] (Aijcc) {$A_{ij,4}$};
	
		\draw[e] (Li) -- (Aic);
		\draw[e] (Li) -- (Lj);
		\draw[e] (Lj) -- (Aijc);
		\draw[e] (Lj) -- (AijA);
		\draw[e] (Lj) -- (Aijx);
		\draw[e] (Lj) -- (Aijcc);
            \end{tikzpicture}}
\end{tabular}
\caption{Computation of the product of an upper triangular matrix with a vector and its tree representation.\label{fig:exampleprogramandtree}}
\end{figure}
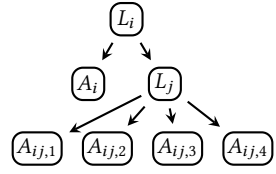

Each loop node $L$ has the following attributes:
\begin{compactitem}
	\item An \emph{iteration domain} $L.\domain$  that captures the values of loop iterators for which the loop is executed.
		The dimensionality of $L.\domain$ depends on the nesting level of the loop node: The root node's domain is one dimensional, and each nesting level adds one dimension, corresponding to the loop iterator of the loop node.
	\item A list of children $L.\children$. Children are either loop nodes or access nodes. Their order defines the order in which the children are to be visited during simulation.
\end{compactitem} 
By construction, the iteration domain of each loop is traversed in lexicographic order.

Each access node $A$ has the following attributes:
\begin{compactitem}
	\item An \emph{iteration domain} $A.\domain$ that captures the loop iterations in which the access should be performed. This is required to model memory accesses that are guarded by a conditional within a loop.
	\item An \emph{access function} $A.\access$ that determines the accessed memory block for each access instance. %
\end{compactitem}

Consider the example program implementing a matrix-vector product computation and its tree representation in Figure~\ref{fig:exampleprogramandtree}. 
Child nodes are sorted from left to right, corresponding to the execution order.

\newcommand{\linearize}{\textit{linearize}}
\newcommand{\start}{\textit{start}}
\newcommand{\block}{\textit{block}}

In our example, the iteration domains are
\begin{align*}
	L_i.\domain & = A_i.\domain = \{(i) \in \mathbb{Z}^1 \mid 0 \leq i < 100\},\\
	L_j.\domain &= \{(i, j) \in \mathbb{Z}^2 \mid 0 \leq i < 100 \wedge i \leq j < 100\}\\
			  &= A_{ij,1}.\domain = A_{ij,2}.\domain = A_{ij,3}.\domain = A_{ij,4}.\domain.
\end{align*} 
As none of the access nodes are guarded by a conditional, their iteration domains are equal to those of their enclosing loops.  
The access functions are
\begin{align*}
 	A_i.\access & = \{(i) \rightarrow \block(\linearize(c[i]))\},\\
 	A_{ij,1}.\access & = \{(i,j) \rightarrow \block(\linearize(c[i]))\},\\
 	A_{ij,2}.\access & = \{(i,j) \rightarrow \block(\linearize(A[i][j]))\},\\
 	A_{ij,3}.\access & = \{(i,j) \rightarrow \block(\linearize(x[j]))\},\\
 	A_{ij,4}.\access & = \{(i,j) \rightarrow \block(\linearize(c[i]))\},
\end{align*}
where $\linearize(\cdot)$ converts an array expression into an expression capturing the accessed memory address. E.g. assuming a row-major layout and an array $A[23][42]$ of 4-byte integers,  
$\linearize(A[i][j]) = \start_A + 42\cdot4 i + 4 j$.
As caches operate at the granularity of memory blocks, $\block$ translates the accessed address into the corresponding memory block, i.e., $\block(x) := \lfloor x/64\rfloor$ assuming a block size of 64 bytes.

\newcommand{\initial}{\textit{initial}}
\newcommand{\final}{\textit{final}}
\newcommand{\vfinal}{\vec{\final}}
\newcommand{\stride}{\vec{\textit{stride}}}
\newcommand{\interval}{\textit{interval}\xspace}

We use \textit{pet}, the \emph{Polyhedral Extraction Tool}~\cite{Verdoolaege2012} to obtain polyhedral representations of SCoPs, which we subsequently transform into the tree representation introduced above.
For convenience during simulation, we also define the following helper functions, which can be implemented using~\isl:
	$L.\initial(\vj) := \lexmin(L.\domain \cap (\{\vj\} \times \mathbb{Z}))$, and %
	$L.\final(\vj) := \lexmax(L.\domain \cap (\{\vj\} \times \mathbb{Z}))$,
which provide the smallest and the largest elements of the iteration domain of $L$ for a fixed assignment of the outer dimensions of the iteration domain.
In addition, $\interval(\vi, \vj) := \{\vk \mid \vi \preceq \vk \prec \vj\}$ captures the set of integers in the interval between $\vi$ and $\vj$. %
Similarly, we may extract the \emph{stride} $L.\stride$ of a loop node, which is the increment of the iteration variable of loop node~$L$.\looseness=-1 %

\newcommand{\this}{\textit{this}}

\begin{algorithm}[t]
\caption{Non-warping cache simulation\label{alg:nonwarping}}
\begin{algorithmic}[1]
\Procedure{LoopNode::Simulate}{$\vj$}
  \State $\vi \gets \this.\initial(\vj)$
  \State $\vfinal \gets \this.\final(\vj)$
  \While{$\vi \preceq \vfinal$}	
    \If{$\vi \in \this.\domain$}
    	\ForAll{$child \in \this.\children$}
    		\State $c \gets child.\Call{Simulate}{\vi}$
	\EndFor
    \EndIf
    \State $\vi \gets \vi + \this.\stride$
  \EndWhile
\EndProcedure
\Procedure{AccessNode::Simulate}{$\vj$}
  \If{$\vj \in \this.\domain$}
	\State $m \gets m+\classifycache(c, \this.\access(\vj))$
	\State $c \gets \updatecache(c, \this.\access(\vj))$
  \EndIf
\EndProcedure
\end{algorithmic}
\end{algorithm}

\section{Non-Warping Cache Simulation of Polyhedral~Programs}

Algorithm~\ref{alg:nonwarping} shows how to perform non-warping cache simulation on top of the tree representation introduced in the previous section.
The algorithm uses two global variables, $c$ and $m$, the current cache state and the current cache miss count.\looseness=-1

To analyze a SCoP, the simulation is initiated by invoking the $\textsc{Simulate}$ procedure of the root node of the tree.
The first parameter, $\vj$, is the state of those loop iterators that are defined in ancestors of a node.
Thus, at the top level, the zero-dimensional tuple $\vj = ()$ can be passed to the procedure. 

For a loop node, the simulator steps through the iteration domain from the initial state $\this.\initial(\vj)$ to the final state $\this.\final(\vj)$.
At each point in the iteration domain the simulation of all child nodes is triggered.
The check in line~5 is required to support guarded statements.

Memory accesses are simulated at access nodes. 
If the current iterator state~$\vj$ is in the access's domain, the cache state is 
updated and the cache miss count is incremented based on the classification of the memory access associated with the current iterator state~$\vj$.

As the SCoP cache simulation may be initiated with any cache state and any cache miss count, the SCoP simulation could be integrated into more general simulation frameworks that apply to non-static control parts of a program.
This also applies to the warping cache simulation that we introduce in the following section. %
However, experimental evaluation of such an integration is outside of the scope of this paper due to the significant required engineering effort.

\section{Warping Cache Simulation of Polyhedral Programs}

We now show how to exploit the data independence of caches in order to speed up cache simulation.
The basic idea is to identify recurring patterns of cache states and memory accesses during the cache simulation and to ``warp'' across these.
In Section~\ref{sec:concretewarping} we introduce the warping theorem that formalizes the above idea.
To efficiently determine candidates for warping, we introduce symbolic cache simulation and a corresponding symbolic warping theorem in Section~\ref{sec:symbolicsimulation}. 
Based on these foundations we finally introduce a warping symbolic cache simulation algorithm in Section~\ref{sec:warpingsimulation}.

\subsection{Concrete Cache Warping}\label{sec:concretewarping}

Warping is based on the following theorem, which follows from Theorem~\ref{thm:dataindependence}: %
\begin{restatable}[Cache warping]{thm}{warping}\label{thm:warping}
Let $c_0, c_1 \in \cachestate$,
	\ifmainpart
\hfill~~\\ 
	\fi
$s_0, \dots, s_n \in \blocks^*$, and $\pi \in \Piindex$, s.t.
\begin{align}
	c_1  = \updatecache(c_0, s_0) = \pi(c_0), \\
	\forall i, 0 \leq i < n: s_{i+1} = \pi(s_i).\label{eq:piseqrelation}
\end{align}
Then:
\begin{align}
	\updatecache(c_1, s_1 \cdot \ldots \cdot s_n) &= \pi^n(c_1), \\
	\classifycache(c_1, s_1 \cdot \ldots \cdot s_n) &= n \cdot \classifycache(c_0, s_0),
\end{align}
where $\cdot$ denotes the concatenation operator on access sequences.
\end{restatable}

In other words, if, during the simulation we arrive at cache state $c_1$ that is equal up to a bijection on the cache contents to an earlier cache state $c_0$, i.e., $c_1 = \pi(c_0)$, and the subsequent memory accesses correspond to those observed between $c_0$ and $c_1$ under the same bijection $\pi$, then warping can be applied, and the final cache state can be computed directly from $c_1$ solely based on the bijection $\pi$.
Depending on the structure of $\pi$, $\pi^n$ can be computed efficiently, e.g. if $\pi$ corresponds to shifting all addresses by a constant.

\begin{example}
Consider again our running example from Figure~\ref{fig:runningexample} and its concrete simulation on a set-associative cache in Figure~\ref{fig:runningexamplesetassociative}.
After reaching cache state $c_6$ and observing that $c_6 = \pi(c_5)$ with $\pi(i) = i+1$, we can apply Theorem~\ref{thm:warping} to obtain cache state $c_{999}$ at the end of the stencil computation as $c_{999} = \pi^{993}(c_6)$.
Also, the number of misses on the remaining 993 iterations of the loop can be determined as $2\cdot 993$.\looseness=-1
\end{example}

Thus at a high level, a warping-based simulation algorithm could proceed as follows:
(i) Simulate cache accesses concretely, until a cache state is obtained that satisfies the conditions of Theorem~\ref{thm:warping}.
(ii) Analyze the ``future'' memory accesses to determine up to which point Theorem~\ref{thm:warping} can be applied.
(iii) Apply warping accordingly and continue at (i).

A naive implementation would compare each cache state to all cache states encountered before. 
This would be highly inefficient.
To more efficiently determine matching cache states, our simulator instead operates on \emph{symbolic cache states}.  %
Symbolic cache states express the concrete cache state in terms of the iterator state.
Whenever the simulation reaches a symbolic cache state that is \emph{equal} to a symbolic cache state encountered before, this implies that the corresponding concrete cache states are related by a bijection; and this bijection can be extracted efficiently from the symbolic cache states.
Equality of symbolic cache states can be detected efficiently via hashing.

\subsection{Symbolic Simulation and Symbolic Warping}\label{sec:symbolicsimulation}

\newcommand{\symb}{\textit{sym-b}}
\newcommand{\syms}{\textit{sym-s}}
\newcommand{\symc}{\textit{sym-c}}

\newcommand{\seq}{\sigma}

To efficiently determine candidate pairs of matching states, we introduce \emph{symbolic cache states}.
In place of concrete memory blocks, symbolic cache sets and symbolic cache states associate cache lines with \emph{symbolic memory blocks}:
\begin{align*}
	\syms \in \symsetstate &= (\lines \rightarrow (\symblocks \cup \{\invalidline\})) \\ & \shoveright{\hspace{2.5cm}\times \policystate,}\\
	\symc \in \symcachestate &= \set \rightarrow \symsetstate.
\end{align*}
Symbolic memory blocks %
 correspond to the access functions of access nodes in our SCoP representation. 
 Thus symbolic memory blocks represent functions that map the state of the loop iterators to concrete memory blocks. 
Due to the restriction to the polyhedral model the expressions used to represent symbolic memory blocks are always of the form~$\lfloor e / c \rfloor$, where $c$ is a constant corresponding to the block size and $e$ is an affine expression in the loop iterators.
In the following we assume an interpretation function $\sem{\cdot}$ that maps symbolic memory blocks to the functions they represent.\looseness=-1

\begin{example}
Consider an access $A[i][j]$. As discussed in Section~\ref{sec:scops}, this access would be associated with an expression representing the function \[\lambda (i,j).\block(\linearize(A[i][j])) = \lfloor(\start_A + 42\cdot4 i + 4 j)/64\rfloor.\]
\end{example}

Symbolic memory blocks, and by extension symbolic cache states, represent different concrete cache states depending on the state of the loop iterators.
Symbolic cache simulation thus maintains a pair of the current symbolic cache state and the current loop iterator.
Such a pair $(\symc, \vi)$ then concretizes to a concrete cache state by replacing each symbolic memory block by the concrete block it represents under~$\vi$:\looseness=-1
\begin{align*}
	\conc(\symc, \vi) &:= \lambda s.\conc_\set(\symc(s), \vi),\\
	\conc_\set(\syms, \vi) &:= (\lambda l.\sem{\syms.m(l)}(\vi), \syms.ps).
\end{align*}
Symbolic cache states can be updated and accesses can be classified so that the following equalities hold: %
\begin{equation}\label{eq:symaccesscorrect}
\ifmainpart
\begin{multlined}
	\conc(\symupdatecache((\symc, \vi), \symb))= \\\shoveright{\updatecache(\conc(\symc, \vi), \sem{\symb}(\vi)),}\\
	\shoveleft{\symclassifycache((\symc, \vi), \symb) =\hspace{2cm}~}\\\shoveright{\hspace{2cm}\classifycache(\conc(\symc, \vi), \sem{\symb}(\vi)).}
\end{multlined}
\else
\begin{split}
	\conc(\symupdatecache((\symc, \vi), \symb)) & = \updatecache(\conc(\symc, \vi), \sem{\symb}(\vi)),\\
	\symclassifycache((\symc, \vi), \symb) & = \classifycache(\conc(\symc, \vi), \sem{\symb}(\vi)).
\end{split}
\fi
\end{equation}
For convenience, $\symupdatecache$ returns both the updated symbolic cache state and the state of the loop iterators.
A constructive definition of $\symupdatecache$ achieving the above equality is given in the \optional{extended version~\cite{arxivVersion}}{appendix}. %
This allows our simulation to operate on symbolic cache states in place of concrete ones.\looseness=-1

\begin{figure}[t]
\begin{center}
\optional{\includegraphics[width=\linewidth]{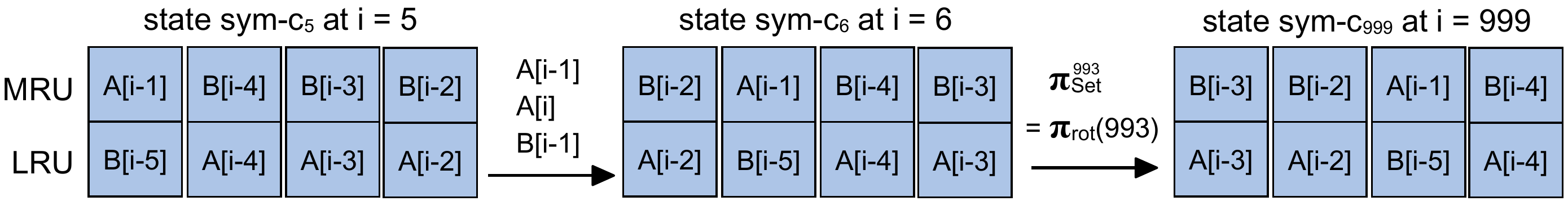}}{\includegraphics[width=0.7\linewidth]{figures/ex-set-sym-warping}}
\end{center}
\caption{Example illustrating symbolic equivalence and symbolic cache warping.\label{fig:symboliccachewarping}}
\end{figure}

As changes to the loop iterators result in a different concretization of symbolic cache states, these have to be adapted upon any increment~$\Delta$ of the loop iterators\footnote{Our implementation determines the updated symbolic cache state only on demand, which significantly increases efficiency.}.
Appropriately adapting the expressions forming a symbolic cache state allows for an update function \symupdatecacheit\ that satisfies the following equality: %
\begin{equation}\label{eq:symdeltacorrect}
	\begin{aligned}
		\symupdatecacheit((\symc, \vi), \Delta) &= (\symc', \vi+\Delta),\\
		\gamma(\symc', \vi+\Delta) &=  \gamma(\symc, \vi).
	\end{aligned}
\end{equation}

\newcommand{\pirot}{\pi_\textit{rot}}

Symbolic cache states are useful to detect opportunities for warping:
\begin{restatable}[Symbolic equivalence of cache states]{thm}{symbolicequivalence}\label{thm:symbolicequivalence}~\\
Let $(\symc_0, \vi_0)$, $(\symc_1, \vi_1) \in \symcachestate \times \mathbb{Z}^n$  and $\pi_\set$ be a bijection on cache sets, s.t. \[\symc_1 = \symc_0 \circ \pi_\set.\] Then there is a $\pi \in \Piindex$, s.t.:
\begin{equation}
	\conc(\symc_1, \vi_1) = \pi(\conc(\symc_0, \vi_0)).
\end{equation}
\end{restatable} 
In other words, if the simulation determines two symbolic cache states that are equal up to a permutation of their cache sets, then their concrete counterparts are also related to each other by a bijection.
To further simplify the search for matches, in our implementation, we are not looking for arbitrary permutations, but only for \emph{rotations}, i.e., permutations of the form $\pirot(c) = \{(i, i+c \bmod s) \mid i \in \set\}$. %

\begin{example}
Consider the symbolic cache states $\symc_5$ and $\symc_6$ in Figure~\ref{fig:symboliccachewarping}, which are the symbolic counterparts to the concrete cache states $c_5$ and $c_6$ from Figure~\ref{fig:runningexamplesetassociative}.
We have $c_6 = \pirot(1)(c_5)$, and thus the two states are symbolically equivalent, which by Theorem~\ref{thm:symbolicequivalence} implies that their concretizations $c_5$ and $c_6$  are related by a bijection.
\end{example}

Having found two symbolic cache states $\symc_1, \symc_2$ that ``match'' does not immediately guarantee that warping can be applied.
This also depends on the accesses between $\symc_1$ and $\symc_2$ and their relation to the subsequent accesses.
The following symbolic warping theorem captures a sufficient condition for warping on symbolic cache states:
\begin{restatable}[Symbolic cache warping]{thm}{symbolicwarping}\label{thm:symbolicwarping}
Let $(\symc_0, \vi_0),\optional{$\\$}{} (\symc_1, \vi_1) \in \symcachestate \times \mathbb{Z}^n$, and $\pi_\set$ be a bijection on cache sets, and $\seq \in \symblocks^*$, such that:
\begin{equation}\label{eq:symupdate}
\begin{aligned}
	\symupdatecache((\symc_0, \vi_0), \seq) &= (\symc_1, \vi_1)\\
		& = (\symc_0 \circ \pi_\set, \vi_1), \\
\end{aligned}
\end{equation}
and let $\pi \in \Piindex$, such that for all $j, 0 \leq j < n$:
\begin{align}
		\conc(\seq, \vi_{j+1}) &= \pi(\conc(\seq, \vi_j)),\label{eq:seqpisymb}\\ 
		\conc(\symc_0 \circ \pi_\set^{j+1}, \vi_{j+1}) &= \pi(\conc(\symc_0 \circ \pi_\set^j, \vi_j)), \label{eq:pioverapprox}  %
\end{align}
with $\vi_j = \vi_0 + j\cdot (\vi_1-\vi_0)$ for $0 \leq j \leq n+1$. 
Then:
\ifmainpart
\begin{multline}
	\updatecache(\conc(\symc_1, \vi_1), \conc(\seq,\vi_1) \cdot \ldots \cdot \conc(\seq,\vi_n)) =
	\\
	\conc(\symc_1 \circ \pi_\set^n, \vi_{n+1}),\label{eq:symbolicwarpingcorrect}
\end{multline}
\begin{multline}
	\classifycache(\conc(\symc_1, \vi_1), \conc(\seq, \vi_1) \cdot \ldots \cdot \conc(\seq, \vi_n)) =
	\\
	n\cdot\symclassifycache((\symc_0, \vi_0), \sigma).\label{eq:symbolicwarpingclassificationcorrect}
\end{multline}
\else
\begin{equation}
	\updatecache(\conc(\symc_1, \vi_1), \conc(\seq,\vi_1) \cdot \ldots \cdot \conc(\seq,\vi_n)) =
	\conc(\symc_1 \circ \pi_\set^n, \vi_{n+1}),\label{eq:symbolicwarpingcorrect}
\end{equation}
\begin{equation}
	\classifycache(\conc(\symc_1, \vi_1), \conc(\seq, \vi_1) \cdot \ldots \cdot \conc(\seq, \vi_n)) =
	n\cdot\symclassifycache((\symc_0, \vi_0), \sigma).\label{eq:symbolicwarpingclassificationcorrect}
\end{equation}
\fi
\end{restatable}
Let's digest this theorem to better understand when and how it is applicable.
Equation (\ref{eq:symupdate}) captures that the matching symbolic cache states need to be equal up to a permutation of their cache sets, and $\symc_1$ needs to be obtained from $\symc_0$ on the symbolic access sequence $\seq$.
Regarding (\ref{eq:symbolicwarpingcorrect}), observe that if all conditions apply, we may warp across $n$ copies of the \emph{same} symbolic access sequence $\seq$. Note, however, that the corresponding $n$ concrete sequences can be different as they concretize under different iterator valuations, e.g. if $\seq$ corresponds to $A[i]; i$++ this would admit warping across potentially many iterations of a loop traversing an array. 
The benefit of $\symc_1 = \symc_0 \circ \pi_\set$ is that warping the symbolic state is achieved by simply applying $\pi_\set^n$ to $\symc_1$.
If $\pi_\set$ is a rotation $\pirot(c)$ with offset $c$, then $\pi_\set^n = \pirot(n\cdot c)$.

Once a match is found, checking applicability requires ensuring that (\ref{eq:seqpisymb}) and (\ref{eq:pioverapprox}) hold. 
Here, (\ref{eq:seqpisymb}) ensures that the concrete sequences corresponding to the $n$~copies of $\seq$ relate to each other under the same bijection~$\pi$.
In practice, (\ref{eq:seqpisymb}) usually holds, but is not guaranteed to in general. Consider e.g. the symbolic sequence corresponding to $A[i]; A[2 \cdot i]; i$++.
For many choices of $i$ and $n$ an appropriate bijection $\pi$ satisfying (\ref{eq:seqpisymb}) exists, but for $i=0$ and $n=2$ this is impossible, as $A[0]; A[2 \cdot 0]$ and $A[1]; A[2\cdot 1]$ cannot be transformed into each other by any bijection.
Finally, condition (\ref{eq:pioverapprox}) ensures that the bijection~$\pi$ is also compatible with the matching symbolic cache states, and their warped versions. %

\begin{example}
In our running example, all conditions of Theorem~\ref{thm:symbolicwarping} hold and thus the final cache state of the loop can be obtained by applying $\pirot^{993}(1) = \pirot(993\cdot 1 \bmod 4) = \pirot(1)$ to $\symc_6$ as shown in the top right corner of Figure~\ref{fig:symboliccachewarping}.\looseness=-1
\end{example}

\newcommand{\vDelta}{\vec{\Delta}}

\begin{algorithm}
\caption{Warping symbolic cache simulation\label{alg:warping}}
\begin{algorithmic}[1]
\Procedure{LoopNode::WarpingSimulate}{$\vj$}
  \State $\vi \gets \this.\initial(\vj)$
  \State $\vfinal \gets \this.\final(\vj)$
  \State $x \gets new~\textit{HashMap}()$
  \While{$\vi \preceq \vfinal$}	
    \If{$x.\textit{contains}(\symc)$}
      \State $(\vi_0, m_0, \pirot) \gets x.get(\symc)$
      \State $\vDelta \gets \vi-\vi_0$ 	\Comment{Match delta}
      \State $n \leftarrow \Call{IterationsToWarp}{\symc, \vi_0, \vi, \vfinal, \vDelta, \pirot}$
        \State $\vi \gets \vi + n\cdot \vDelta$	\Comment{Warp $n\cdot \vDelta$ iterations}\label{line:warpingstarts}
        \State $\symc \gets \symc \circ \pirot^n$						
        \State $m \gets m + n \cdot (m-m_0)$ 						\label{line:warpingends}
    \EndIf
    \State $x.\textit{put}(\symc, (\vi, m))$
    \If{$\neg x.\textit{contains}(\symc) \vee n = 0$}	\Comment{Could not warp}	\label{line:ordinarystarts}
    	\ForAll{$child \in \this.\children$}
    		\State $child.\Call{WarpingSimulate}{\vi}$
	\EndFor
        \State $(\symc, \vi) \gets \symupdatecacheit(\symc, \this.\stride)$			\label{line:ordinaryends}
    \EndIf
  \EndWhile
\EndProcedure

\Procedure{AccessNode::WarpingSimulate}{$\vj$}
  \If{$\vj \in \this.\domain$}
	\State $m \gets m+\symclassifycache(\symc, (\this.\access, \vj)))$
	\State $(\symc, \vi) \gets \symupdatecache(\symc, (\this.\access, \vj)))$
  \EndIf
\EndProcedure 
\end{algorithmic}
\end{algorithm}

\begin{figure}
\begin{algorithmic}[1]
\Procedure{IterationsToWarp}{$\symc, \vi_0, \vi_1, \vfinal, \vDelta, \pirot$}
  \State $\vi_{f_c} \gets \Call{FurthestByDomains}{\vi_0, \vi_1, \vfinal, \vDelta}$
  \State $\vi_{f_a} \gets \Call{FurthestByOverlap}{\vi_0, \vfinal}$
  \State $\vi_f \gets \lexmin(\vi_{f_a}, \vi_{f_c})$
  \If{$\Call{CacheAgrees}{\symc, \vi_0, \vi_1, \vi_f, \vDelta, \pirot}$}
    \State \Return $\lexmax \{n \mid \vi_1 + n\cdot \vDelta \prec \vi_f\}$  
  \EndIf
  \State \Return $0$
\EndProcedure 
\end{algorithmic}

\begin{algorithmic}[1]
\Procedure{FurthestByDomains}{$\vi_0, \vi_1, \vfinal, \vDelta$}
  \State $I_m \gets \interval(\vi_0, \vi_1)$ \Comment{Match interval}
  \State $I_w \gets \interval(\vi_1, \vfinal)$ \Comment{Max. warp interval}
  \State $C \gets \emptyset$ \Comment{Conflict set}
  \ForAll{\textit{AccessNode} $a \in \this.\children^*$}  %
   \State $C_a \gets \{ \vi_c \mid \neg(\vi_c \in (a.\domain \cap I_w) \Leftrightarrow\optional{$ \\ ~\hfill\hfill\hfill~$}{}(\vi_0 + ((\vi_c - \vi_1) \bmod \vDelta)) \in (a.\domain \cap I_m))\}$
    \State $C \gets C \cup C_a$ 
  \EndFor
   \If{$C = \emptyset$}  \Return $\vfinal$  
  \EndIf
  \State \Return $\lexmin(C)$  %
\EndProcedure
\end{algorithmic}

\begin{algorithmic}[1]
\Procedure{FurthestByOverlap}{$\vi_0, \vfinal$}
  \State $I \gets \interval(\vi_0, \vfinal)$ \Comment{Access interval}
  \State $C \gets \emptyset$ \Comment{Conflict set}
  \ForAll{\textit{AccessNode} $a, b \in \this.\children^*$} %
        \If{$a.access$ and $b.access$ have the same coefficients}
   		 \State continue  
 	 \EndIf
  	\State $C_{a, b} \gets \{ \vi \mid \exists \vj_a \in (a.\domain \cap I): \exists \vj_b \in (B.\domain \cap I): \optional{$\\ ~\hfill\hfill\hfill~$}{}a.access(\vj_a) = b.access(\vj_b)$
		  $\land~\vj_a \preceq \vi \land \vj_b \preceq \vi\}$ 
    \State $C \gets C \cup C_{a, b}$
  \EndFor
     \If{$C = \emptyset$} \Return $\vfinal$  
  \EndIf
  \State \Return $\lexmin(C)$ 
\EndProcedure
\end{algorithmic}
\begin{algorithmic}[1]
\Procedure{CacheAgrees}{$\symc, \vi_0, \vi_1, \vi_f, \vDelta, \pirot$}
  \State $\pi \gets \Call{ConstructAccessMapping}{\vi_0, \vi_f, \vDelta}$
  \State $c_0 \gets \conc(\symc, \vi_0)$
  \State $c_1 \gets \conc(\symc, \vi_1)$
  \ForAll{\textit{set} $s$, \textit{line} $l$} 
  	\State $b_0 \gets c_0(s).m(l)$
  	\State $b_1 \gets c_1(\pirot(s)).m(l)$
  	\If{$b_0 \in \pi_{\domain} \land \pi(b_0) \neq b_1$}
  	\Return \textit{false}
	\EndIf
	\If{$b_1 \in \pi_{\textit{ran}} \land \pi^{-1}(b_1) \neq b_0 $}
  	\Return \textit{false}
	\EndIf
  \EndFor
  \State \Return \textit{true}
\EndProcedure

\Procedure{ConstructAccessMapping}{$\vi_0, \vi_f, \vDelta$}
  \State $I \gets \interval(\vi_0, \vi_f)$ \Comment{Access interval}
  \State $\pi \gets \emptyset$
  \ForAll{\textit{AccessNode} $a \in \this.\children^*$}
  	\State $\pi_a \gets \{ (b_1) \rightarrow (b_2) \mid \exists \vj \in (a.\domain \cap I) \land\optional{$ \\ ~\hfill\hfill\hfill~$} b_1 = a.\access(\vj) \land b_2 = a.\access(\vj + \vDelta)\}$
    \State $\pi \gets \pi \cup \pi_a$%
  \EndFor
  \State \Return $\pi$
\EndProcedure
\end{algorithmic}
\end{figure}

\subsection{Warping Symbolic Cache Simulation}\label{sec:warpingsimulation}

Let us now explain the warping symbolic cache simulation algorithm in Algorithm~\ref{alg:warping}, which is based upon Theorem~\ref{thm:symbolicwarping} and applies polyhedral techniques to ensure applicability of the theorem for warping.

The algorithm differs from Algorithm~\ref{alg:nonwarping} in two ways: (i) it applies symbolic rather than concrete cache simulation, and (ii) it applies warping.
   
To find matching symbolic cache states, each loop node maintains a separate hash map in which it stores symbolic cache states reached since the last change to an iterator of an enclosing loop.
Thus warping is only attempted across different iterations of a loop while staying in the same iteration of all enclosing loops.
This is a deliberate choice that slightly reduces the ability to warp but greatly reduces ``spurious'' matches that do not result in actual warping opportunities.

The hash value of a symbolic cache state is determined based on the symbolic memory blocks in the cache.
To identify ``rotating'' matches, the hash computation does not begin at a fixed set, but rather starts at the most-recently-accessed cache set and from there on cycles around the cache sets. 
Thus, when a match is determined, the difference between the indexes of the most-recently-accessed cache sets of the matching cache states determines the relative rotation.\looseness=-1

If a matching cache state is found, \textsc{IterationsToWarp} determines the number of iterations %
 to warp across.
Lines~\ref{line:warpingstarts} to~\ref{line:warpingends} carry out the warping and update the number of misses~$m$.
In the worst case $n = 0$ and the simulation needs to proceed via ordinary symbolic cache simulation in lines~\ref{line:ordinarystarts} to~\ref{line:ordinaryends}, which is also applied if there is no match.  

The procedure \textsc{IterationsToWarp} relies on three sub-procedures to determine how many iterations to warp across:

	(i) \textsc{FurthestByDomains} determines up to which iteration the future symbolic memory accesses are identical to the symbolic memory accesses in the match interval. This is determined by separately considering the domains of every access node that is a descendant of the warping loop node.
	The set~$C_a$ is constructed such that it contains all iterator valuations that conflict with the corresponding iteration in the match interval, i.e., either the corresponding iteration was present in the match interval but is missing in $\vi_c$ or vice versa.
	Based on Theorem~\ref{thm:symbolicwarping} warping is limited to repetitions of the same symbolic access sequence, and so warping across such conflicts is impossible.
	Thus the earliest conflict is determined using $\lexmin(C)$.
	
	(ii) To satisfy (\ref{eq:seqpisymb}), there needs to be a single bijection~$\pi$ that applies to all accesses in $\conc(\sigma, \vi_i)$ for all $i$.
		These accesses may stem from different access nodes, which may each depend differently on loop iterators.
		Consider for example two access nodes with access expressions $A[i+50]$ and $A[i+j]$.
		Warping in a loop node with loop iterator $j$, where loop iterator $i$ corresponds to an enclosing loop, implies that the bijection $\pi$ must map $A[i+50]$ to $A[i+50]$ for the first access node, and $A[i+j]$ to $A[i+j+1]$ for the second access node (assuming the matching cache states differ by 1 in $j$).
		If $j$ may obtain the value $50$ this would yield conflicting requirements on the joint bijection $\pi$.
		Thus, for any two access nodes with conflicting coefficients on the warped loop iterator, \textsc{FurthestByOverlap} determines the maximal loop iteration for which the ranges of the iterators do not overlap.
		
	(iii) Finally, \textsc{CacheAgrees} checks whether the relation induced by the access sequences is compatible with the matching symbolic cache states. To this end, \textsc{Construct\-AccessMapping} incrementally constructs the minimal required relation to satisfy (\ref{eq:seqpisymb}) and \textsc{CacheAgrees} checks whether this conflicts with the induced relation between the concretizations of the matches, which corresponds to (\ref{eq:pioverapprox}).\looseness=-1

\subsubsection*{Multi-level Caches}
For simplicity, we have presented warping cache simulation for single-level caches.
However, it is equally applicable to multi-level caches, and our implementation currently supports two-level non-inclusive non-exclusive cache hierarchies.
The simulation then operates on pairs of symbolic L1 and L2 states and updates the L1 and the L2 state upon each memory access according to the inclusion policy. 
Warping is performed whenever both the L1 and the L2 cache states satisfy the conditions of Theorem~4.

\section{Experimental Evaluation}\label{sec:evaluation}

In our evaluation we aim to answer the following questions:
1.~What are the benefits of warping cache simulation in terms of simulation performance?
2.~How does warping cache simulation compare with analytical approaches such as HayStack and PolyCache?
3.~How strong is the influence of different replacement policies on cache performance?

\subsection{Experimental Setup}

We implemented our approach as a cache simulation tool which takes as input the cache parameters and a C program, and outputs cache access and miss counts. 
We use \emph{pet-0.11} (Polyhedral Extraction Tool)~\cite{Verdoolaege2012} to extract the polyhedral model from the C source and \emph{isl-0.22} (Integer Set Library)~\cite{Verdoolaege2010} to perform operations on integer sets.
We plan to release the source code of our tool as open source.
\looseness=-1

We evaluate our cache simulation tool on \emph{PolyBench 4.2.1-beta}~\cite{Pouchet2012}, a benchmark suite of numerical computations implemented as SCoPs. 
PolyBench benchmarks are configurable with different problem sizes. %
The experiments that we present here are for the large (L) and extra large (XL) problem sizes; the two largest ones.

We run our experiments single threaded using only one core on a test system with Intel Core i9-10980XE (Cascade Lake) processors.
Unless stated otherwise, the cache simulation assumes the cache configuration found in the test system itself:
Each core has an 8-way set-associative \SI{32}{KiB} L1 cache with Pseudo-LRU replacement policy and a 16-way set-associative \SI{1}{MiB} L2 cache with Quad-age LRU replacement policy both with a block size of 64 bytes.
Both L1 and L2 caches are write-back write-allocate and the inclusion policy between L1 and L2 is non-inclusive non-exclusive~\cite{Solihin2015}.

A replication package for our experiments is available~\cite{Morelli22artifact}.

\subsection{Warping vs Non-Warping Simulation}

\begin{figure}
    \centering
	\includegraphics[width=\linewidth]
	{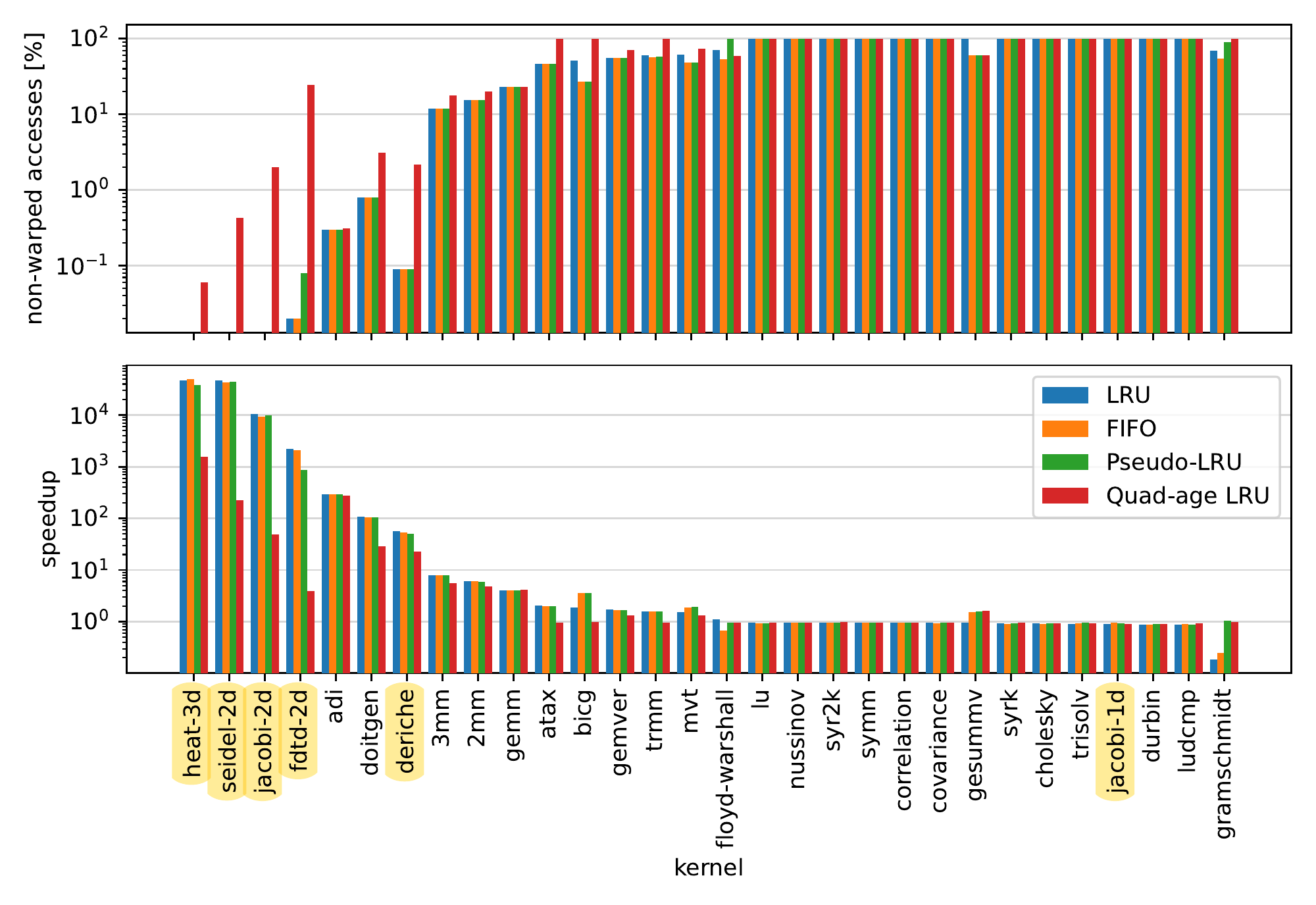}
    \caption{Speedup of L1 warping simulation compared to non-warping simulation (bottom) and share of non-warped accesses (top) for LRU, FIFO, Pseudo-LRU, and Quad-age LRU and problem size~L.}
    \label{fig:nonwarping-vs-warping-l}
\end{figure}

\subsubsection*{Warping vs Non-warping Simulation}
We first simulate the L1 cache of the test system for problem size~L.
To investigate the effect of the replacement policy on the warping performance, in addition to the Pseudo-LRU policy of the test system, we also simulate LRU, FIFO, and Quad-age LRU.

Figure~\ref{fig:nonwarping-vs-warping-l} shows for each benchmark the speedup of warping simulation compared to the non-warping simulation (bottom) and the share of non-warped accesses (top).
The reported times correspond to the time spent executing the implementations of Algorithms~\ref{alg:nonwarping} and~\ref{alg:warping}, i.e., they \emph{do not} include the overhead of extracting the internal representation of the benchmarks via \isl. 
This overhead, which is identical for warping and non-warping simulation, lies between 62 and 245~ms depending on the benchmark and is thus dominated by the simulation time for most benchmarks.

The first observation is that the speedup is roughly inversely proportional to the share of non-warped accesses.
For, e.g. \textit{adi}, about 0.3\% of all accesses cannot be warped and we observe a speedup of about $300$x.

The stencil kernels \textit{adi}, \textit{fdtd-2d}, \textit{heat-3d}, \textit{jacobi-2d}, and \textit{seidel-2d} exhibit large speedups.
Stencils have \emph{uniformly generated references}~\cite{Gannon1988,Wolf1991}, and thus give rise to recurring patterns in the cache if there are enough accesses relative to the cache size.
As we discussed earlier, warping aims to exploit these patterns to accelerate the simulation.
The consistent speedups for the stencil kernels show that warping simulation is indeed able to achieve this.
The \textit{jacobi-1d} kernel does not benefit from warping since its working set is too small to fill the cache.%

While there are many kernels that benefit from warping, there are others that do not.
We observed that there were no (or very few) symbolically equivalent cache states during the simulation of these kernels, and thus, no (or very few) opportunities for warping.
As we show later, some of these kernels benefit from warping when simulating a different cache.
However, for the current cache configuration, we conclude that warping does not decrease the simulation times of these kernels.

Overall, the differences between the replacement policies are fairly small, with LRU, Pseudo-LRU, and FIFO often exhibiting similar speedups.
Quad-age LRU is scan- and thrash-resistant~\cite{jaleel10}, which may result in ``old'' memory blocks remaining in the cache, while scanning through new ones, which in some cases results in a greater number of classic simulation steps before detecting warping opportunities.

\subsubsection*{Impact of Problem Size}

\newcommand{\maxXY}{300000000}
\newcommand{\markScale}{0.60}
\newcommand{\arrowthickness}{very thin}
\pgfplotsset{width=8cm,height=8cm,compat=1.12}
\newcommand{\scatterPlot}{
	\begin{tikzpicture}[scale=0.80, transform shape]
	\begin{loglogaxis}[
		ymin=0.35, ymax=\maxXY,
		xmin=0.35, xmax=\maxXY,	
		xlabel={non-warping simulation time [ms]},
		ylabel={warping simulation time [ms]},
		legend style={
			at={(1.05,0.0)},
			anchor=south west}
	]
	
\draw [->, \arrowthickness] (765412, 128278) -- (5966037, 196618);
\draw [->, \arrowthickness] (1305724, 162639) -- (10024582, 227014);
\draw [->, \arrowthickness] (1379432, 4676) -- (10762495, 8702);
\draw [->, \arrowthickness] (3071, 1521) -- (3012, 1376);
\draw [->, \arrowthickness] (3037, 839) -- (2961, 931);
\draw [->, \arrowthickness] (543167, 585904) -- (4327575, 4620025);
\draw [->, \arrowthickness] (500584, 527947) -- (6414185, 6584657);
\draw [->, \arrowthickness] (495775, 523916) -- (6384455, 6545568);
\draw [->, \arrowthickness] (20430, 407) -- (75205, 602);
\draw [->, \arrowthickness] (229390, 2200) -- (1644325, 900428);
\draw [->, \arrowthickness] (1380, 1542) -- (5710, 6304);
\draw [->, \arrowthickness] (823816, 948) -- (7150964, 764);
\draw [->, \arrowthickness] (14274661, 15082643) -- (113336160, 110452312);
\draw [->, \arrowthickness] (540335, 134403) -- (4879883, 114265);
\draw [->, \arrowthickness] (5652, 3342) -- (22116, 5373);
\draw [->, \arrowthickness] (1349, 862) -- (6037, 1671);
\draw [->, \arrowthickness] (725791, 689268) -- (8485346, 4627716);
\draw [->, \arrowthickness] (1880621, 49) -- (19265994, 45);
\draw [->, \arrowthickness] (711, 761) -- (2961, 17);
\draw [->, \arrowthickness] (1018061, 104) -- (9180922, 49);
\draw [->, \arrowthickness] (1205005, 1292786) -- (9482130, 9927745);
\draw [->, \arrowthickness] (747788, 840564) -- (5647026, 6154328);
\draw [->, \arrowthickness] (3286, 1693) -- (12887, 3147);
\draw [->, \arrowthickness] (1952530, 2037800) -- (20916049, 21621739);
\draw [->, \arrowthickness] (1814735, 41) -- (14653249, 40);
\draw [->, \arrowthickness] (434825, 457642) -- (4991234, 5139174);
\draw [->, \arrowthickness] (515735, 540075) -- (4844064, 4993371);
\draw [->, \arrowthickness] (338812, 363655) -- (3130388, 3287268);
\draw [->, \arrowthickness] (802, 833) -- (3076, 3233);
\draw [->, \arrowthickness] (303493, 192718) -- (3157475, 721410);

	\addplot[
	scatter,
	only marks,
	point meta=explicit symbolic,
	scatter/classes={
		2mm_L={mark=*, draw=blue!80!black, fill=blue, scale=\markScale},%
		2mm_XL={mark=*, draw=green!80!black, fill=green, scale=\markScale},%
		3mm_L={mark=*, draw=blue!80!black, fill=blue, scale=\markScale},%
		3mm_XL={mark=*, draw=green!80!black, fill=green, scale=\markScale},%
		adi_L={mark=*, draw=blue!80!black, fill=blue, scale=\markScale},%
		adi_XL={mark=*, draw=green!80!black, fill=green, scale=\markScale},%
		atax_L={mark=*, draw=blue!80!black, fill=blue, scale=\markScale},%
		atax_XL={mark=*, draw=green!80!black, fill=green, scale=\markScale},%
		bicg_L={mark=*, draw=blue!80!black, fill=blue, scale=\markScale},%
		bicg_XL={mark=*, draw=green!80!black, fill=green, scale=\markScale},%
		cholesky_L={mark=*, draw=blue!80!black, fill=blue, scale=\markScale},%
		cholesky_XL={mark=*, draw=green!80!black, fill=green, scale=\markScale},%
		correlation_L={mark=*, draw=blue!80!black, fill=blue, scale=\markScale},%
		correlation_XL={mark=*, draw=green!80!black, fill=green, scale=\markScale},%
		covariance_L={mark=*, draw=blue!80!black, fill=blue, scale=\markScale},%
		covariance_XL={mark=*, draw=green!80!black, fill=green, scale=\markScale},%
		deriche_L={mark=*, draw=blue!80!black, fill=blue, scale=\markScale},%
		deriche_XL={mark=*, draw=green!80!black, fill=green, scale=\markScale},%
		doitgen_L={mark=*, draw=blue!80!black, fill=blue, scale=\markScale},%
		doitgen_XL={mark=*, draw=green!80!black, fill=green, scale=\markScale},%
		durbin_L={mark=*, draw=blue!80!black, fill=blue, scale=\markScale},%
		durbin_XL={mark=*, draw=green!80!black, fill=green, scale=\markScale},%
		fdtd-2d_L={mark=*, draw=blue!80!black, fill=blue, scale=\markScale},%
		fdtd-2d_XL={mark=*, draw=green!80!black, fill=green, scale=\markScale},%
		floyd-warshall_L={mark=*, draw=blue!80!black, fill=blue, scale=\markScale},%
		floyd-warshall_XL={mark=*, draw=green!80!black, fill=green, scale=\markScale},%
		gemm_L={mark=*, draw=blue!80!black, fill=blue, scale=\markScale},%
		gemm_XL={mark=*, draw=green!80!black, fill=green, scale=\markScale},%
		gemver_L={mark=*, draw=blue!80!black, fill=blue, scale=\markScale},%
		gemver_XL={mark=*, draw=green!80!black, fill=green, scale=\markScale},%
		gesummv_L={mark=*, draw=blue!80!black, fill=blue, scale=\markScale},%
		gesummv_XL={mark=*, draw=green!80!black, fill=green, scale=\markScale},%
		gramschmidt_L={mark=*, draw=blue!80!black, fill=blue, scale=\markScale},%
		gramschmidt_XL={mark=*, draw=green!80!black, fill=green, scale=\markScale},%
		heat-3d_L={mark=*, draw=blue!80!black, fill=blue, scale=\markScale},%
		heat-3d_XL={mark=*, draw=green!80!black, fill=green, scale=\markScale},%
		jacobi-1d_L={mark=*, draw=blue!80!black, fill=blue, scale=\markScale},%
		jacobi-1d_XL={mark=*, draw=green!80!black, fill=green, scale=\markScale},%
		jacobi-2d_L={mark=*, draw=blue!80!black, fill=blue, scale=\markScale},%
		jacobi-2d_XL={mark=*, draw=green!80!black, fill=green, scale=\markScale},%
		lu_L={mark=*, draw=blue!80!black, fill=blue, scale=\markScale},%
		lu_XL={mark=*, draw=green!80!black, fill=green, scale=\markScale},%
		ludcmp_L={mark=*, draw=blue!80!black, fill=blue, scale=\markScale},%
		ludcmp_XL={mark=*, draw=green!80!black, fill=green, scale=\markScale},%
		mvt_L={mark=*, draw=blue!80!black, fill=blue, scale=\markScale},%
		mvt_XL={mark=*, draw=green!80!black, fill=green, scale=\markScale},%
		nussinov_L={mark=*, draw=blue!80!black, fill=blue, scale=\markScale},%
		nussinov_XL={mark=*, draw=green!80!black, fill=green, scale=\markScale},%
		seidel-2d_L={mark=*, draw=blue!80!black, fill=blue, scale=\markScale},%
		seidel-2d_XL={mark=*, draw=green!80!black, fill=green, scale=\markScale},%
		symm_L={mark=*, draw=blue!80!black, fill=blue, scale=\markScale},%
		symm_XL={mark=*, draw=green!80!black, fill=green, scale=\markScale},%
		syr2k_L={mark=*, draw=blue!80!black, fill=blue, scale=\markScale},%
		syr2k_XL={mark=*, draw=green!80!black, fill=green, scale=\markScale},%
		syrk_L={mark=*, draw=blue!80!black, fill=blue, scale=\markScale},%
		syrk_XL={mark=*, draw=green!80!black, fill=green, scale=\markScale},%
		trisolv_L={mark=*, draw=blue!80!black, fill=blue, scale=\markScale},%
		trisolv_XL={mark=*, draw=green!80!black, fill=green, scale=\markScale},%
		trmm_L={mark=*, draw=blue!80!black, fill=blue, scale=\markScale},%
		trmm_XL={mark=*, draw=green!80!black, fill=green, scale=\markScale}	
	},
	]
	table[x=nowarp, y=warp, meta=kernel] {scatter-plot-data.txt};
	\addplot[black, domain=0.01:\maxXY, samples=2]{x};
	\addplot[gray, domain=0.01:\maxXY, samples=2]{10*x};
	\addplot[gray, domain=0.01:\maxXY, samples=2]{100*x};
	\addplot[gray, domain=0.01:\maxXY, samples=2]{1000*x};
	\addplot[gray, domain=0.01:\maxXY, samples=2]{10000*x};
	\addplot[gray, domain=0.01:\maxXY, samples=2]{100000*x};
	\addplot[gray, domain=0.01:\maxXY, samples=2]{1000000*x};
	\addplot[gray, domain=0.01:\maxXY, samples=2]{10000000*x};
	\addplot[gray, domain=0.01:\maxXY, samples=2]{100000000*x};
	\addplot[gray, domain=0.01:\maxXY, samples=2]{1000000000*x};
	\addplot[gray, domain=0.01:\maxXY, samples=2]{x/10};
	\addplot[gray, domain=0.01:\maxXY, samples=2]{x/100};
	\addplot[gray, domain=0.01:\maxXY, samples=2]{x/1000};
	\addplot[gray, domain=0.01:\maxXY, samples=2]{x/10000};
	\addplot[gray, domain=0.01:\maxXY, samples=2]{x/100000};
	\addplot[gray, domain=0.01:\maxXY, samples=2]{x/1000000};
	\addplot[gray, domain=0.01:\maxXY, samples=2]{x/10000000};
	\addplot[gray, domain=0.01:\maxXY, samples=2]{x/100000000};
	\addplot[gray, domain=0.01:\maxXY, samples=2]{x/1000000000};
	\addplot[gray, domain=0.01:\maxXY, samples=2]{x/10000000000};

	\legend{L,XL}

	\end{loglogaxis}
	\end{tikzpicture}
}

\begin{figure}
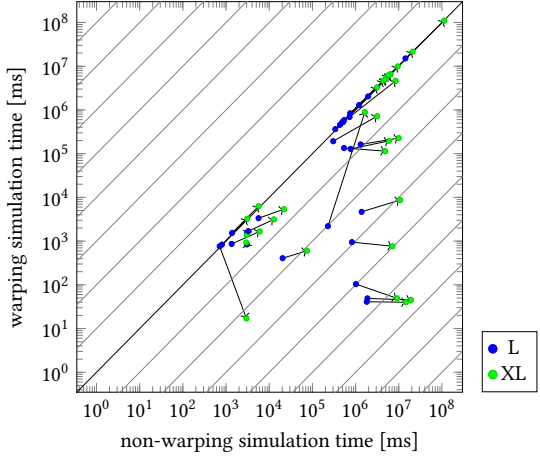

\centering
\scatterPlot
\hskip 2pt
\caption{L1 warping and non-warping simulation times for problem sizes L and XL.} %
\label{fig:scatter-plot-l-xl}
\end{figure}

Figure~\ref{fig:scatter-plot-l-xl} shows the change in warping and non-warping L1 simulation times between problem sizes L and XL for the configuration of the test system, i.e., with Pseudo-LRU replacement.
We can see that for many benchmarks the warping simulation times are not proportional to the number of memory accesses while the non-warping simulation times are.
On the other hand, there are also benchmarks whose warping simulation times change considerably between L and XL problem sizes.
One interesting observation is that there are benchmarks whose simulation is faster with the XL problem size.
This is unintuitive at first but it can happen when the simulator is able to warp across more accesses.
Consider the simulation of a loop that has 10 iterations left.
A matching cache state from 20 iterations ago cannot be used to warp across the last 10 iterations, as 10 is not a multiple of 20. 
However, for a larger problem size with e.g. 1000 iterations left, the analysis could warp to the end of the loop.
This can have a considerable effect on the simulation time, especially when it applies at the outermost level of a deeply nested loop.

\subsection{Warping Simulation vs Analytical Cache Modeling Approaches}

We compare the performance of warping simulation to the analytical models PolyCache~\cite{Bao2018} and HayStack~\cite{Gysi2019}.

\subsubsection*{Warping Simulation vs HayStack}

\begin{figure}[htpb]
  \centering
  \optional{\includegraphics[width=\linewidth]
	{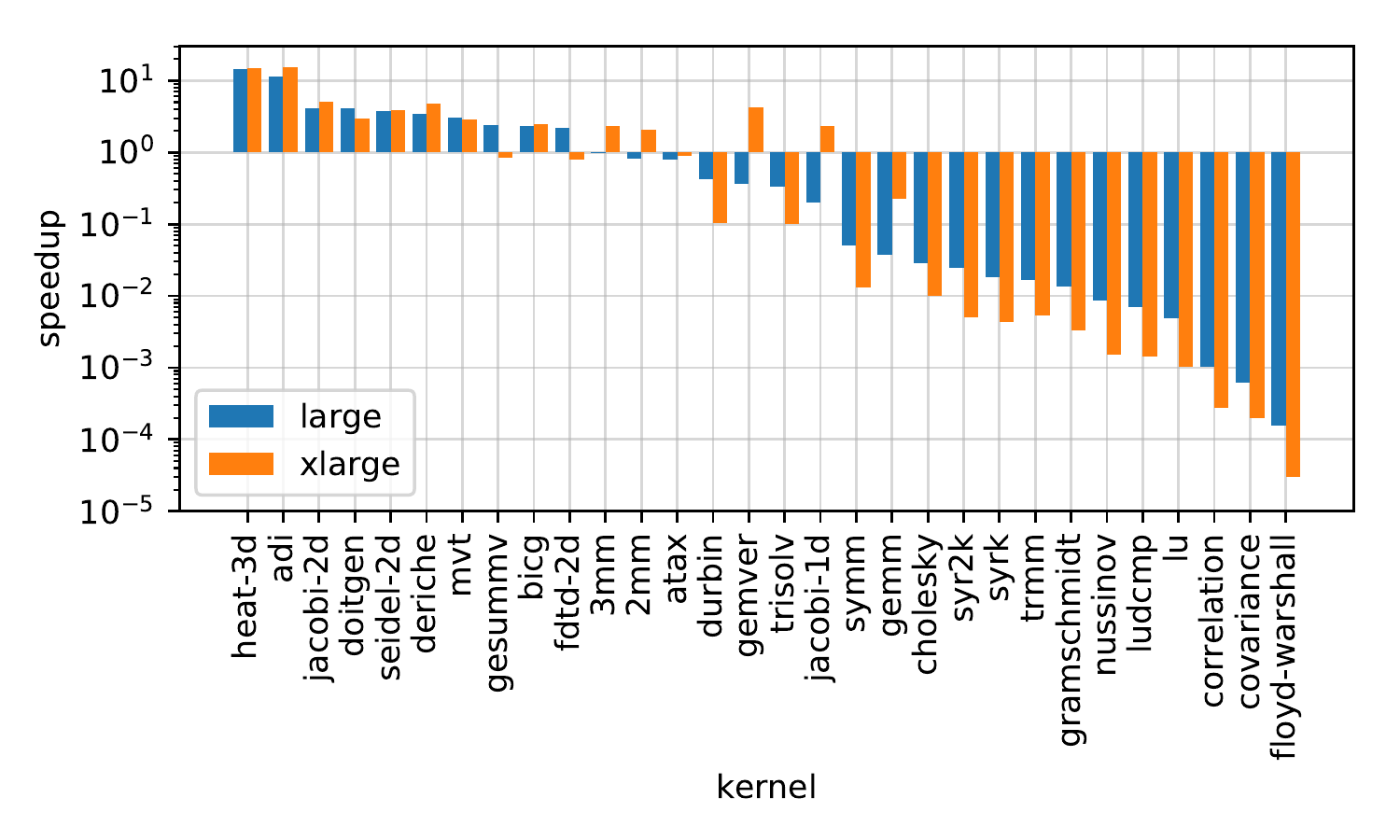}}{\includegraphics[width=0.7\linewidth]
	{plots/haystack-vs-warping-l1-large-xlarge-speedup.pdf}}  
  \caption{Speedup of L1 warping simulation compared to HayStack for problem sizes L and XL.}
  \label{fig:haystack-vs-warping}
\end{figure}

We compare warping simulation to the analytical cache model HayStack~\cite{Gysi2019}.
HayStack provides a replication package~\cite{artifacthaystack}, which we use to replicate their experimental systems on our test system.
We simulate the fully-associative LRU version of the L1 cache of the test system as HayStack can only model fully-associative caches with LRU replacement.
As the HayStack experiments \emph{do} include the overhead of extracting the internal representation of the benchmarks, we include this overhead also for warping simulation in this comparison.  

Figure~\ref{fig:haystack-vs-warping} shows the speedup of warping simulation compared to HayStack for each kernel for problem sizes L and XL when the L1 cache is simulated.
We can see that HayStack is faster than warping cache simulation on most, but not all benchmarks. %
In particular, warping cache simulation outperforms HayStack on most stencil benchmarks.
On the XL problem size, HayStack extends its lead on those benchmarks for which warping simulation is not effective, i.e., where its runtime is proportional to the number of accesses.
Conversely, warping remains faster on stencil benchmarks on the XL problem size.\looseness=-1

\subsubsection*{Warping Simulation vs PolyCache}

We compare warping simulation to the analytical model PolyCache~\cite{Bao2018} using the published results, as no replication package is available.
For this purpose, we run warping simulation using the same PolyBench problem size (L) and cache configuration as PolyCache: a two-level cache with \SI{32}{KiB} 4-way set-associative L1 and \SI{256}{KiB} 4-way set-associative L2 caches.
Both caches employ LRU replacement, write-allocate write-back write policy, and 64-byte cache blocks.

\begin{figure}
\centering
	\optional{\includegraphics[width=\linewidth]
	{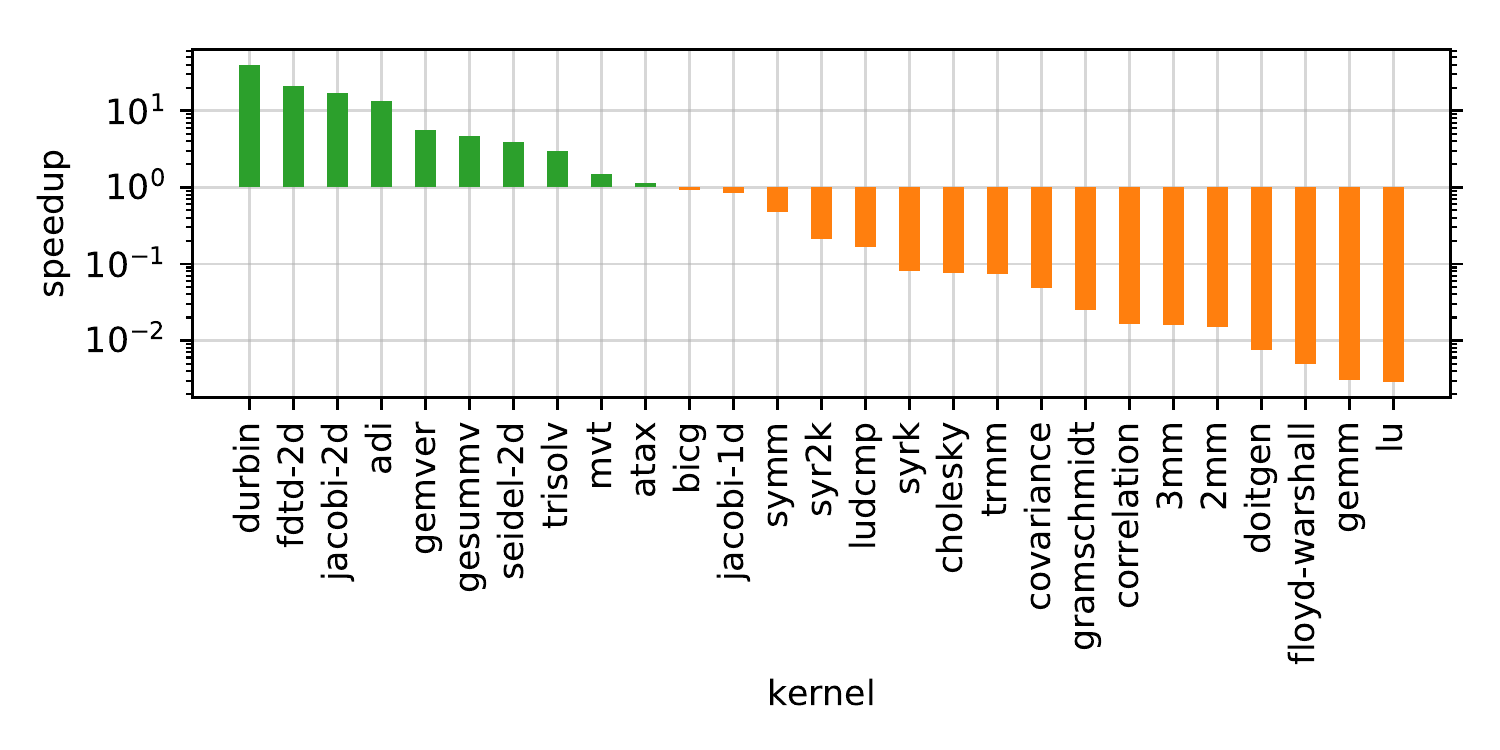}}{\includegraphics[width=0.7\linewidth]
	{plots/polycache-vs-warping-l1-l2-large-speedup.pdf}}
    \caption{Speedup of L1-L2 warping simulation compared to PolyCache for problem size L.}
    \label{fig:polycache-vs-warping}
\end{figure}

Figure~\ref{fig:polycache-vs-warping} shows the speedup of warping simulation compared to PolyCache for each benchmark.
On the average, PolyCache outperforms warping cache simulation, but the relative performance varies greatly across the set of benchmarks.
Note that this experiment is missing some of the PolyBench kernels as they are not included in the PolyCache results. 
We cannot correct for differences in the hardware on which the simulation is carried out. 
The simulation times for our tool include all overheads as in the comparison with HayStack. 
It is unclear from the documentation, which overheads are included in the PolyCache results. 
In contrast to our single-threaded implementation, PolyCache analyzes each of the 128 cache sets in a separate process.
Thus, the single-thread performance of PolyBench would be expected to be about $128$x slower.\looseness=-1

\subsection{Influence of Parameters on Cache Performance}

\subsubsection*{Influence of the Replacement Policy}
\begin{figure}
    \centering
	\includegraphics[width=\linewidth]
	{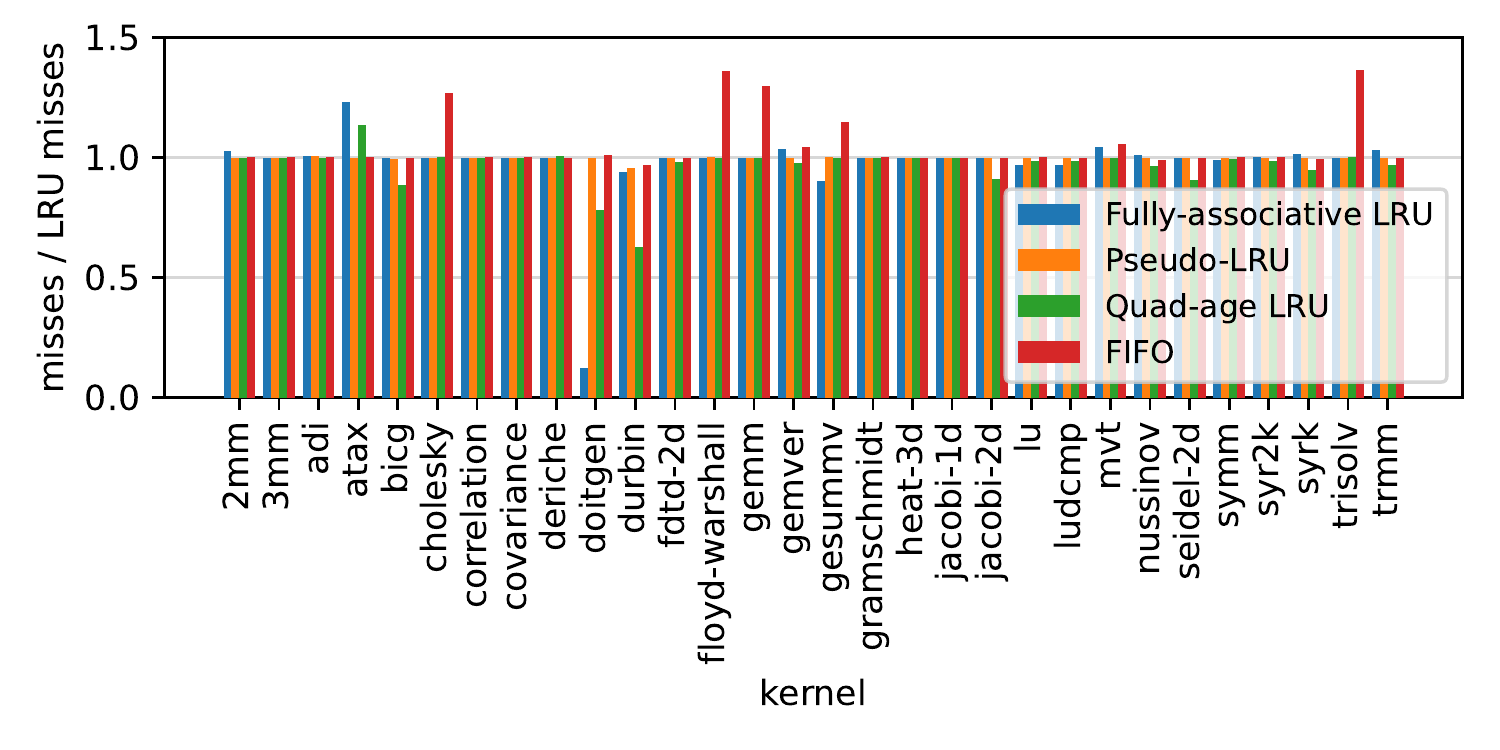}
    \optional{\vspace{-6mm}}{}
    \caption{Number of misses relative to set-associative LRU.} 
       \label{fig:repl-pol-miss}
    \optional{\vspace{-1mm}}{}
\end{figure}

To determine the influence of the replacement policy, %
we simulate each benchmark, again using problem size L, under the four replacement policies LRU, FIFO, Pseudo-LRU, and Quad-age LRU on a \SI{32}{KiB} 8-way set-associative L1 cache with 64-byte cache blocks.
The results of this experiment are depicted in Figure~\ref{fig:repl-pol-miss}.
For most PolyBench benchmarks, cache performance does not vary dramatically depending on the replacement policy, but there are notable exceptions.
In particular, on a number of benchmarks, e.g. \emph{durbin} and \emph{doitgen}, Quad-age LRU achieves significant improvements over LRU, while FIFO sometimes incurs significantly more misses.
This demonstrates that accurately modeling the replacement policy can be important.

\subsubsection*{Comparison with Measurements on Actual Hardware}
We also evaluate the accuracy of cache simulation by comparing the number of cache misses predicted by the various simulators to PAPI-C~\cite{Terpstra2010} measurements on a real system.
We compile the PolyBench~\cite{Pouchet2012} kernels with -O2 GCC optimization level and PAPI-C support to measure the cache misses on the test system.
Note that PolyBench flushes the cache before executing each kernel.
To minimize measurement errors, we repeat them 10 times and take the median.
We also disable cache prefetching in our test system.  

We compare the cache misses simulated by Dinero~IV, warping simulation, and HayStack to the measured misses on problem size L. 
Dinero IV simulates a set-associative LRU cache whereas HayStack models a same-size fully-associative LRU cache.
Warping simulation simulates the system cache as it is, set-associative with Pseudo-LRU replacement policy.
Note that as memory accesses, Dinero~IV considers both array and scalar accesses while warping simulation and HayStack consider only array accesses. %

\begin{figure}
    \centering
	\includegraphics[width=\linewidth]
	{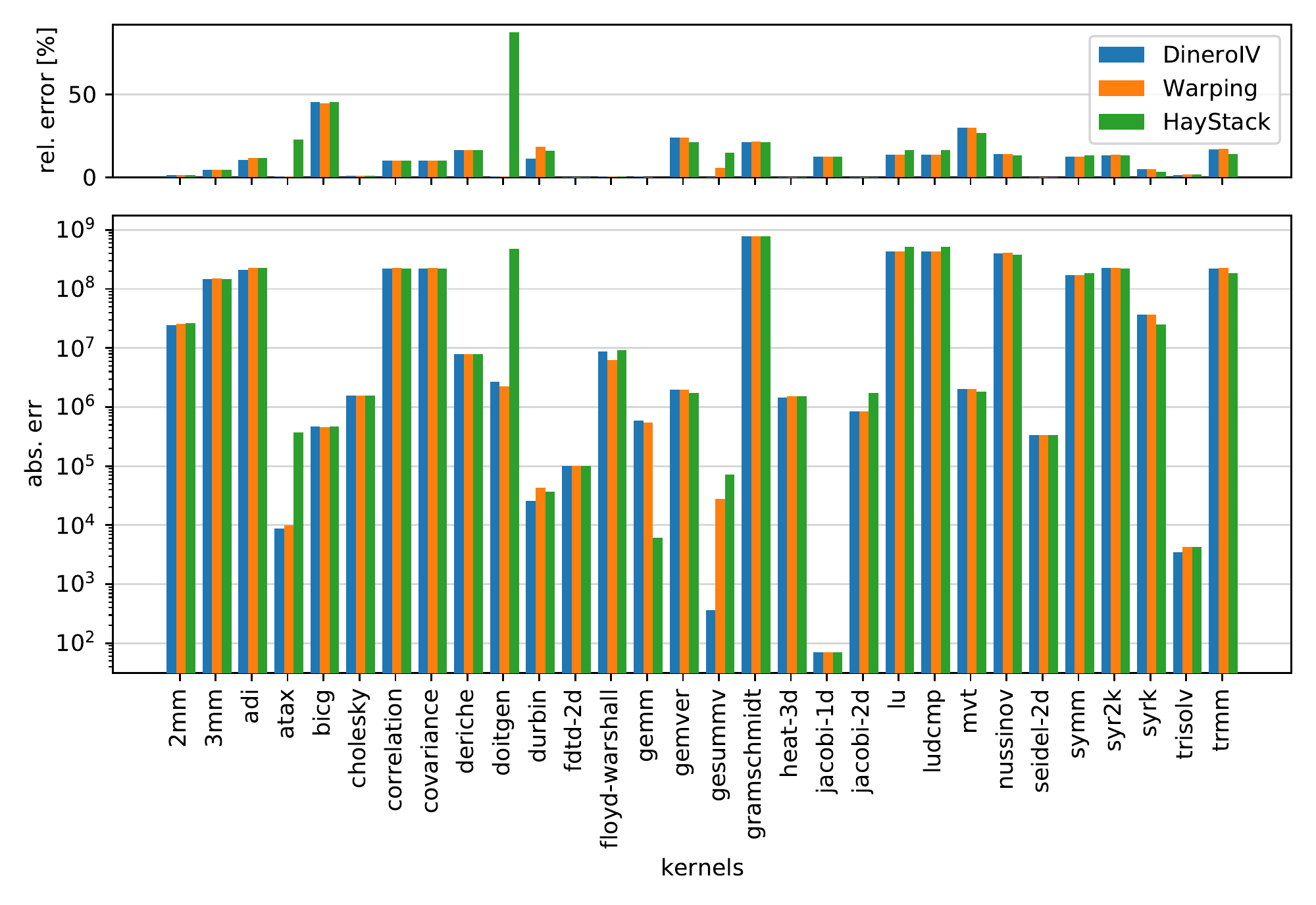}
    \optional{\vspace{-6mm}}{}
    \caption{Accuracy relative to measurements on the actual hardware using PAPI.}
    \label{fig:papi-accuracy-l1}
    \optional{\vspace{-1mm}}{}
\end{figure}

When comparing the measured misses to the simulated or modeled cache misses, we consider two main metrics:
\begin{compactenum}
    \item Absolute error: the absolute value of the difference between actual and predicted number of cache misses. 
    \item Relative error: absolute error divided by the  actual number of misses.
\end{compactenum}
The results of this experiment are depicted in Figure~\ref{fig:papi-accuracy-l1}.
For most benchmarks, all analytical approaches are similarly accurate, with some exceptions, e.g. on \emph{atax} and \emph{doitgen}, HayStack is significantly less accurate, due to its modeling of a fully-associative cache.
In the \optional{extended version~\cite{arxivVersion}}{appendix} we present additional experimental results for other problem sizes, where more significant accuracy differences are observed.
The main takeaway though is that other aspects of modern microarchitectures, such as memory reordering and speculative execution, have a strong influence on cache performance not captured by any of the present approaches.
Future work will have to further investigate this discrepancy.\looseness=-1

\section{Known Limitations}
Our implementation is currently limited to caches with \emph{modulo placement}, i.e. to caches in which a memory block $b$'s cache set is determined by $\cacheindex(b) = b \bmod s$, where $s$ is the cache's number of sets.\looseness=-1

Modulo placement is common in L1 and L2 caches.
However, shared L3 caches are often ``sliced''. 
Sliced L3 caches are split into equally-sized slices, where each slice is physically located at one of the processor's cores.
To distribute data evenly across the slices, pseudo-random hash functions are used to determine a memory block's slice~\cite{Hund13,Maurice15,Yarom15}.\looseness=-1

Such pseudo-random hash functions do not violate data independence.
However, their pseudo-random nature will make it less likely to detect rotating matches that are compatible with the induced access mappings.
Possibly a more flexible mechanism to detect matches may remedy this, but this is left for future work.

Similarly, following Equation~\ref{eq:defupdatecache} and Theorem~\ref{thm:dataindependence}, our implementation currently only applies to caches that treat all cache sets the same and each cache set independently of other cache sets. 
Several mechanisms have been described in the literature that conflict with these assumptions.
Qureshi et al.~\cite{Qureshi07} propose \emph{set dueling}, which has subsequently been implemented in Intel processors~\cite{jaleel10,Wong10}. 
In set dueling, a small number of dedicated leader cache sets are used to test two or more competing replacement policies. 
The remaining follower sets use the policy that performed best on the leader sets in the previous epoch.
To reduce cache misses, Rol\'{a}n et al.~\cite{Rolan09} propose \emph{set-balancing caches} in which cache sets are paired up statically or even dynamically, moving cache lines from highly saturated to less saturated ones.
Skewed-associative caches~\cite{Seznec93} and the ZCache~\cite{Sanchez10} even do away with the notion of cache sets altogether and share a cache's lines more flexibly.
All these cache designs fundamentally \emph{do} satisfy data independence; but they exhibit fewer symmetries and may thus offer fewer warping opportunities.
It remains future work to explore the potential of warping for such designs.

\section{Related Work}

\subsubsection*{Traditional Cache Simulators} 
Cache simulators such as Dinero~IV~\cite{Edler1999} and CASPER~\cite{Iyer2003} simulate the cache behavior of a program by explicitly iterating over the memory access traces that are generated by the program.
This approach applies to arbitrary programs and can model modern memory hierarchies precisely, including inclusive, non-inclusive non-exclusive, and exclusive cache hierarchies as well as sophisticated cache replacement policies such as Pseudo-LRU~\cite{al04} and Quad-age LRU~\cite{jaleel10,jahagirdar12}, which are employed in recent real-world microarchitectures~\cite{Vila2020,Abel2020}.
The main drawback is that the simulation cost is {proportional to the number of memory accesses} that the simulated program performs.\looseness=-1

\subsubsection*{Analytical Cache Models} 
There is a long history of analytical cache models~\cite{Ghosh1997,Ghosh1999,Chatterjee2001,Vera2002,Vera2004,Cascaval2003,Beyls2005,Bao2018,Gysi2019}.
Seminal work by Ghosh et al.~\cite{Ghosh1997,Ghosh1999} introduces cache miss equations (CMEs), systems of linear Diophantine equations, that capture the set of cache misses of a loop nest. 
Their approach builds upon Wolf and Lam's~\cite{Wolf1991} characterization of data reuse in loop nests via reuse vectors, and its classification into self-spatial, self-temporal, group-spatial, and group-temporal reuse.
CMEs capture when these different types of reuse do not result in cache hits in single-level set-associative caches with LRU replacement.
An inherent limitation of reuse vectors is their inability to accurately capture reuse in programs with conditional statements and between different references that are not uniformly generated.

Cascaval and Padua~\cite{Cascaval2003} present an exact approach to compute stack histograms~\cite{Mattson1970} of programs at compile time.
Stack histograms immediately reveal the number of cache misses under single-level fully-associative LRU caches for any given cache size, thus allowing to gauge the impact of different cache sizes on a program's cache performance.
Their work is limited to the same class of programs as Ghosh et al.'s CMEs~\cite{Ghosh1997,Ghosh1999}, but a larger class of programs can be modeled approximately in this framework, sacrificing accuracy. 
 We note that, when applied to LRU caches, our approach could similarly be extended to compute stack histograms rather than the number of misses for a fixed cache size.
 Based on prior work by Smith and Hill~\cite{Smith1978,Hill1989}, Cascaval and Padua~\cite{Cascaval2003} also show how to approximate the number of misses of a program under a set-associative cache based on its stack histogram.

Vera and Xue~\cite{Vera2002} extend the applicability of CMEs to a larger class of programs involving conditional statements, multiple loop nests, and subroutines by transforming such projects into a more restricted normal form. 
Some of these transformations, however, approximate the original program's behavior, rendering the analysis inexact. 
 Vera et al.~\cite{Vera2004} apply sampling techniques to efficiently estimate the number of solutions of CMEs with statistical guarantees, rather than counting the number of solutions exactly.
 Similarly, Chen et al.~\cite{Chen2018} introduce static sampling of reuse times, where a random subset of a program's memory accesses is sampled and subsequently analyzed.
 Varying the number of sampled accesses then allows to trade off performance and accuracy of the predictions.

Chatterjee et al.~\cite{Chatterjee2001} introduce a compositional characterization of the cache behavior of polyhedral programs~\cite{Feautrier1991} via Presburger formulas~\cite{Haase2018} for single-level set-associative caches with LRU replacement.
Their approach takes into account the initial cache state and distinguishes interior misses and boundary misses, which allows to analyze sequential programs in a compositional manner.  
At the time of publication, the approach did not scale to realistic levels of associativity. 

More recent work by Bao et al.~\cite{Bao2018} introduces PolyCache, a tool applicable to polyhedral programs, just like Chatterjee et al.~\cite{Chatterjee2001} and our work.
By analytically characterizing the sequence of cache misses at a given cache level, it can incrementally handle write-allocate non-inclusive non-exclusive~\cite{Solihin2015} \emph{multi-level} set-associative caches with LRU replacement. Non-write allocate caches are handled approximately.
Similarly to \cite{Chatterjee2001} their method constructs an integer set consisting of a program's cache misses. 
Then, \isl's~\cite{Verdoolaege2010} implementation of Barvinok's algorithm is used to compute the integer set's size, and thus the number of cache misses.
As we have shown in the experimental evaluation, PolyCache outperforms warping cache simulation on the average, but the relative performance varies greatly across the set of benchmarks, and unlike our work the approach is not applicable to replacement policies other than LRU and it handles non-write allocate caches approximately.

Gysi et al.~\cite{Gysi2019} present HayStack, the most scalable analytical cache analysis approach to date.
Their approach applies to polyhedral programs, but it is limited to fully-associative LRU caches and inclusive hierarchies of such caches, which do not require to model the interaction between different cache levels.
HayStack performs symbolic counting twice: first, a Presburger relation is constructed that relates each access A to its ``conflict set'', i.e., the set of distinct memory blocks accessed between the most-recent access to the memory block accessed by A.
The size of this conflict set, is the access's stack distance.
In a fully-associative LRU cache, an access results in a cache miss if and only if its stack distance is greater than the cache's associativity.
In the first step, the stack distances of all accesses are determined by symbolic counting of the conflict sets.
This step is inspired by prior work of Beyls and D'Hollander~\cite{Beyls2005}.
In the second step, the number of accesses with a stack distance greater than the cache's associativity is determined, again by symbolic counting. 
This second step is challenging, as the stack distances are generally non-affine. 
Non-affine terms are eliminated by a partially explicit enumeration before applying symbolic counting.
Thus, as our work, HayStack can in fact be characterized as a hybrid approach. 
HayStack is ``analytical first'' and resorts to explicit simulation enumeration as a last resort.
In contrast, our approach is ``simulation first'' and applies analytical reasoning for warping.
The experimental evaluation demonstrates that HayStack is more scalable than our approach on the average, but it is limited to fully-associative LRU caches.\looseness=-1

Both PolyCache and HayStack are currently limited to LRU caches.
Can this restriction be lifted easily?
We believe not, as both approaches deeply rely on the following favorable property of LRU that does not hold under any other policy: 
To classify a memory access A as a hit or a miss, it is sufficient to consider the size of A's ``conflict set'', i.e., the set of distinct memory blocks accessed between the most-recent access to the memory block accessed by A.
Under other policies it is necessary to take into account (a) the cache state prior to the most-recent access to the block accessed by A, and (b) the order of the conflicting memory accesses.
``Simulation-first'' approaches such as ours naturally account for such state and ordering effects.\looseness=-1

\subsubsection*{Analytical Program Representations}
The polyhedral model \cite{Feautrier1991,Feautrier1992,Benabderrahmane2010} is the basis of our work and many of the recent analytical cache models~\cite{Chatterjee2001,Bao2018,Gysi2019}.
One of its original applications and that of related techniques~\cite{Karp1967,Lamport1974} was to capture data dependencies to facilitate the automatic generation of parallel schedules.
More recently~\cite{Wolf1991,Bacon1994,Bondhugula2008,Verdoolaege2013} it has also been applied to generate schedules that exhibit more locality.\looseness=-1

\subsubsection*{Static Cache Analysis}
Static cache analyses~\cite{Alt1996,Grund2009,Grund2010,Grund2010b,Guan13,Chattopadhyay2013,Guan14,Griffin2014,Touzeau2017,Touzeau2019} bound a program's cache behavior for all possible program executions.
This is different from cache simulation, which applies to a particular program execution.
For polyhedral programs, whose data-access behavior is input independent these goals align.
However, existing static cache analyses do not classify each dynamic memory access separately, but rather collectively classify sets of accesses, e.g. all accesses corresponding to a memory reference. %
Due to their coarse classification granularity even exact static analyses~\cite{Chattopadhyay2013,Touzeau2017,Touzeau2019} overapproximate a polyhedral program's cache misses. %
Existing cache analyses are generally also handcrafted to particular replacement policies in contrast to this work, which applies to arbitrary policies that satisfy the data-independence property.
Most cache analyses are tailored to LRU~\cite{Alt1996,Chattopadhyay2013,Touzeau2017,Touzeau2019}, while some work is dedicated to FIFO~\cite{Grund2009,Grund2010,Guan13}, NMRU~\cite{Guan14}, and Pseudo-LRU~\cite{Grund2010b,Griffin2014}. 
To our knowledge, there is no static cache analysis for Quad-age LRU. %
Monniaux and Touzeau~\cite{Monniaux2019} study the complexity of static cache analysis for different replacement policies.
Cache analysis problems are NP-complete under LRU while they are PSPACE-complete under FIFO, NMRU, and Pseudo-LRU.

\subsubsection*{Acceleration Techniques}
Warping cache simulation bears some resemblance of \emph{acceleration} techniques~\cite{Boigelot1994,Comon1998,Annichini2001,Bardin2003,Su2004,Leroux2007,Gawlitza2009}.
Such techniques accelerate the computation of the reachable set of states of a given model by computing the exact effect of iterating through a control cycle in the model, where the control cycle to iterate is determined dynamically during the analysis.
In contrast, the cycles that warping cache simulation iterates across are generated by the composition of a polyhedral program and a cache model, and thus the cycle is not present explicitly in the input to the analyzer.
Nevertheless, further exploring the connection to acceleration techniques might be fruitful.

\section{Conclusions and Future Work}

We have introduced warping cache simulation and demonstrated its benefits experimentally:
Warping may speed up simulation by several orders of magnitude.
In contrast to existing analytical approaches, warping cache simulation may accurately model replacement policies of real-world cache architectures.
 A natural target for future work is to apply warping to efficiently simulate modern speculative out-of-order processors core and branch prediction mechanisms and their interaction with the cache, which promises to increase the accuracy of the predictions w.r.t. real hardware.

\begin{acks}                            %
  We thank the anonymous reviewers, our shepherd Gabriel Rodr\'{i}guez, and Valentin Touzeau
  	for their constructive feedback, which has helped improve this paper.
  This project has received funding from the European Research Council (ERC) under the European Union’s Horizon 2020 research and innovation programme (grant agreement No. 101020415). 
\end{acks}

\bibliography{biblio}

\ifmainpart
\else

\appendix
\section{Supplementary Material}

\subsection{Proofs}
 
\dataindependence*
\begin{proof}
Note that by definition of $\pi_\set$, we have 
\begin{equation}
	\pi_\set(\cacheindex(b)) = \cacheindex(\pi(b)).\label{eq:indexequality}
\end{equation}
\[
\begin{array}{rcl}
	&& \pi(\updatecache(c,b))\\
& \overset{(\ref{eq:defupdatecache})}{=} & \pi(c[\cacheindex(b) \mapsto \updateset(c(\cacheindex(b)),b)])\\
 &\overset{(\ref{eq:defpicachestate})}{=} & \lambda i. \pi(c[\cacheindex(b) \mapsto \updateset(c(\cacheindex(b)),b)](\pi_\set^{-1}(i))) \\
& = & \lambda i.\pi(c(\pi_\set^{-1}(i)))[\pi_\set(\cacheindex(b)) \mapsto \pi(\updateset(c(\cacheindex(b)),b))]\\
 &\overset{(\ref{eq:defpicachestate})}{=} & \pi(c)[\pi_\set(\cacheindex(b)) \mapsto \pi(\updateset(c(\cacheindex(b)),b))]\\
 &\overset{(\ref{eq:dataindependenceset})}= & \pi(c)[\pi_\set(\cacheindex(b)) \mapsto \updateset(\pi(c(\cacheindex(b))),\pi(b))]\\
& = & \pi(c)[\pi_\set(\cacheindex(b)) \mapsto \updateset(\pi(c(\pi_\set^{-1}(\pi_\set(\cacheindex(b))))),\pi(b))]\\
& \overset{(\ref{eq:indexequality})}= & \pi(c)[\cacheindex(\pi(b)) \mapsto \updateset(\pi(c(\pi_\set^{-1}(\cacheindex(\pi(b))))),\pi(b))]\\
& \overset{(\ref{eq:defpicachestate})}{=} & \pi(c)[\cacheindex(\pi(b)) \mapsto \updateset(\pi(c)(\cacheindex(\pi(b))),\pi(b))]\\
 &\overset{(\ref{eq:defupdatecache})}{=} & \updatecache(\pi(c), \pi(b))
\end{array}
\]

\[ 
\begin{array}{rcl}
	&& \classifycache(c,b))\\
& \overset{(\ref{eq:defclassifycache})}{=} & \classifyset(c(\cacheindex(b)), b)\\

& = & \classifyset(\lambda i.\pi(c(\pi_\set^{-1}(i)))(\pi_\set(\cacheindex(b))), \pi(b))\\
& \overset{(\ref{eq:defpicachestate})}{=} & \classifyset(\pi(c)(\pi_\set(\cacheindex(b))), \pi(b))\\
& \overset{(\ref{eq:indexequality})}{=} & \classifyset(\pi(c)(\cacheindex(\pi(b))), \pi(b))\\

 &\overset{(\ref{eq:defclassifycache})}{=} & \classifycache(\pi(c),\pi(b)))
\end{array}
\]
\end{proof}

\let\originalleft\left
\let\originalright\right
\renewcommand{\left}{\mathopen{}\mathclose\bgroup\originalleft}
\renewcommand{\right}{\aftergroup\egroup\originalright}

\warping*
\begin{proof}

In our proof, we will make use of the fact that
\begin{equation}\label{eq:helperpis}
	\forall i, 1 \leq i \leq n: s_i = \pi^i(s_0),
\end{equation}
which follows directly by induction from (\ref{eq:piseqrelation}).

Proof of $\forall i, 1 \leq i \leq n: \updatecache(c_1, s_1 \cdot \ldots \cdot s_i) = \pi^i(c_1)$:

(Base case, $i=0$):\\
We have $\updatecache(c_1, \epsilon) = c_1 = \pi^0(c_1)$.

(Induction step, $i \rightarrow i+1$):\\
\[
\begin{array}{cl}
	   & \updatecache(c_1, s_1 \cdot \ldots \cdot s_{i+1})\\
	\overset{\textit{Def. }\updatecache}= & \updatecache(\updatecache(c_1, s_1 \cdot \ldots \cdot s_i), s_{i+1})\\
	\overset{\textit{Inductive Hypothesis}}= & \updatecache(\pi^i(c_1), s_{i+1})\\
	\overset{(\ref{eq:helperpis})}=	&	 \updatecache(\pi^i(c_1), \pi^{i+1}(s_0))\\
	\overset{c_1=\pi(c_0)}=	&	 \updatecache(\pi^{i+1}(c_0), \pi^{i+1}(s_0))\\
	\overset{\textit{Theorem \ref{thm:dataindependence}}}=	& \pi^{i+1}(\updatecache(c_0, s_0))\\
	\overset{c_1=\updatecache(c_0, s_0)}=	& \pi^{i+1}(c_1)
\end{array}
\]

Proof of $\forall i, 1 \leq i \leq n: \classifycache(c_1, s_1 \cdot \ldots \cdot s_i) = i \cdot 
\classifycache(c_0, s_0)$:

(Base case, $i=0$):\\
We have $\classifycache(c_1, \epsilon) = 0 = 0 \cdot \classifycache(c_0, s_0)$

(Induction step, $i \rightarrow i+1$):\\
\[
\begin{array}{cl}
	   & \classifycache(c_1, s_1 \cdot \ldots \cdot s_{i+1})\\
	\overset{\textit{Def. }\classifycache}= & \classifycache(c_1, s_1 \cdot \ldots \cdot s_i) + \classifycache(\updatecache(c_1, s_1 \cdot \ldots \cdot s_i), s_{i+1})\\
	\overset{\textit{Inductive Hypothesis}}= & i\cdot \classifycache(c_0, s_0) + \classifycache(\updatecache(c_1, s_1 \cdot \ldots \cdot s_i), s_{i+1})\\
	\overset{\textit{Previous proof}}=	&	 i\cdot \classifycache(c_0, s_0) + \classifycache(\pi^i(c_1), s_{i+1})\\
	\overset{(\ref{eq:helperpis})}=	&	 i\cdot \classifycache(c_0, s_0) + \classifycache(\pi^i(c_1), \pi^{i+1}(s_0))\\
	\overset{c_1=\pi(c_0)}=& i\cdot \classifycache(c_0, s_0) + \classifycache(\pi^{i+1}(c_0), \pi^{i+1}(s_0))\\
	\overset{\textit{Theorem \ref{thm:dataindependence}}}=	&  i\cdot \classifycache(c_0, s_0) + \classifycache(c_0, s_0)\\
	=	& (i+1) \cdot \classifycache(c_0, s_0)
\end{array}
\]
\end{proof}

\newcommand{\mypi}{\pi_0}

\symbolicequivalence*
\begin{proof}

Let
\begin{align*}
	\mypi :=~ & \{(\sem{\symc_0(s).m(l)}(\vi_0), \sem{(\symc_1 \circ \pi_\set^{-1})(s).m(l)}(\vi_1)) \mid s \in \set, l \in \lines\}\notag\\
			 	=~& \{(\sem{\symc_0(s).m(l)}(\vi_0), \sem{\symc_0(s).m(l)}(\vi_1)) \mid s \in \set, l \in \lines\}%
\end{align*}

Notice that by construction $\conc(\symc_1, \vi_1) = \mypi(\conc(\symc_0, \vi_0))$.
As discussed in the previous proof, no memory block may be simultaneously cached in multiple cache lines, and so $\mypi$ is injective, and can thus be extended to a bijection $\pi$, such that $\conc(\symc_1, \vi_1) = \pi(\conc(\symc_0, \vi_0))$.
\end{proof}

\symbolicwarping*
\begin{proof}
Relying on Theorem~\ref{thm:warping} we will first prove $\forall j. 0 \leq j \leq n$:
\begin{align*}
	 \updatecache(\conc(\symc_1, \vi_1), \conc(\seq,\vi_1) \cdot \ldots \cdot \conc(\seq,\vi_j)) &= \pi^j(\conc(\symc_1, \vi_1)),\\
	 \classifycache(\conc(\symc_1, \vi_1), \conc(\seq,\vi_1) \cdot \ldots \cdot \conc(\seq,\vi_j)) &= j \cdot \classifycache(\conc(\symc_0, \vi_0), \conc(\seq, \vi_0)).
\end{align*}
It follows from (\ref{eq:symupdate}) and (\ref{eq:symaccesscorrect}) that $\conc(\symc_1, \vi_1) = \updatecache(\conc(\symc_0, \vi_0), \conc(\sigma, \vi_0))$.\\
Further, from (\ref{eq:pioverapprox}) with $j=0$ and (\ref{eq:symbolicwarpingcorrect}), we have that $\conc(\symc_1, \vi_1) = \pi(\conc(\symc_0, \vi_0))$.
Thus \[\conc(\symc_1, \vi_1) = \updatecache(\conc(\symc_0, \vi_0), \conc(\sigma, \vi_0)) = \pi(\conc(\symc_0, \vi_0)).\]
Let $c_0 = \conc(\symc_0, \vi_0)$ and $c_1 = \updatecache(\conc(\symc_0, \vi_0), \conc(\sigma, \vi_0)) = \updatecache(c_0, s_0) = \pi(c_0)$ with $s_0, \dots, s_n = \conc(\sigma,\vi_0), \dots, \conc(\sigma, \vi_n)$.
Thanks to (\ref{eq:seqpisymb}), we can  apply Theorem~\ref{thm:warping} to $c_0, c_1$ and $s_0, 
\dots, s_n$ and conclude $\forall j. 0 \leq j \leq n$:
\begin{align}
	 \updatecache(\conc(\symc_1, \vi_1), \conc(\seq,\vi_1) \cdot \ldots \cdot \conc(\seq,\vi_j)) &= \pi^j(c_1),\label{eq:updateconcrete}\\
	 \classifycache(\conc(\symc_1, \vi_1), \conc(\seq,\vi_1) \cdot \ldots \cdot \conc(\seq,\vi_j)) &= j \cdot \classifycache(c_0, s_0).\label{eq:classifyconcrete}
\end{align}
By (\ref{eq:symaccesscorrect}), we also have
\begin{align*}
	\classifycache(c_0, s_0) = \classifycache(\conc(\symc_0, \vi_0), \conc(\seq, \vi_0)) &= \symclassifycache((\symc_0, \vi_0), \seq),
\end{align*}
which allows us to conclude (\ref{eq:symbolicwarpingclassificationcorrect}) from (\ref{eq:classifyconcrete}).

To finish the proof of (\ref{eq:symbolicwarpingcorrect}) based on (\ref{eq:updateconcrete}), we will now show that 
\begin{equation*}
	\pi^j(c_1)  = \conc(\symc_1 \circ \pi_\set^j, \vi_{j+1}) \textit{ for all } j, 0\leq j \leq n.
\end{equation*}
Our proof is by induction over $j$.
 
(Induction base, $j = 0$): 
Trivially, $\pi^0(c_1) = \pi^0(\conc(\symc_1, \vi_1)) = \conc(\symc_1 \circ \pi_\set^0, \vi_{0+1})$.

(Induction step, $j \rightarrow j+1$):
\[\begin{tabular}{cl}
	& $\pi^{j+1}(c_1)$\\
$ \overset{\textit{Ind. Hypothesis}}=$&$ \pi(\conc(\symc_1 \circ \pi_\set^j, \vi_{j+1}))$\\
$ \overset{(\ref{eq:pioverapprox})}=$&$ \conc(\symc_1 \circ \pi_\set^{j+1}, \vi_{j+2}))$
\end{tabular}\]

\end{proof}

\subsection{Data Independence of Cache Hierarchies}
    
Memory hierarchies of modern multi-core processors contain multiple cache levels.
Typically, L1 and L2 caches are private to a core, while the L3 cache is shared among the processor's cores.

In the following we model only the private L1 and L2 levels, whose behavior does not depend on co-running tasks on other cores. 
Simulating the behavior of a shared L3 cache would require workloads encompassing co-running tasks.

The state of a two-level cache is captured by a pair capturing the state of the L1 and the L2 cache, respectively:
\begin{equation*}
	c=(c_{L1}, c_{L2}) \in \hierarchycachestate = \cachestate_{L1} \times \cachestate_{L2}
\end{equation*}
Often, L1 and L2 caches of the same processor implement different replacement policies and have different associativities and numbers of sets~\cite{Abel2020,Vila2020}. 
Thus, here and in the following, we use subscripts $1$ and $2$ to refer to the functions and parameters of the L1 and L2 caches.

There are different \emph{inclusion policies}~\cite{Solihin2015}, which determine whether the contents of the L1 cache are included in the contents of the L2 cache, or not.
We model the update of a \emph{non-inclusive non-exclusive} cache hierarchy in the following, but note that \emph{inclusive} and \emph{exclusive} cache hierarchies can be captured in a similar manner:
\begin{equation}\label{eq:defupdatecachehierarchy}
	\updatehierarchicalcache((c_{L1}, c_{L2}),b) := \left(\updatecache_{L1}(c_{L1},b),\right. \left.\begin{cases}
											c_{L2}	&: \classifycache(c_{L1},b)=\textit{true}\\
											\updatecache_{L2}(c_{L2},b)	& : \classifycache(c_{L1},b)=\textit{false}
									           \end{cases}\right)
\end{equation}

The number of sets of the L2 cache is usually a multiple of the number of sets of the L1 cache.
Thus, ${\PiindexTwo} \subset {\PiindexOne}$, where $\PiindexOne$ and $\PiindexTwo$ refer to the variants of the set of bijections $\Piindex$ for the L1 and the L2 caches, respectively.
Under this assumption, we can prove the following corollary of Theorem~\ref{thm:dataindependence}, where $\pi((c_{L1}, c_{L2})) := (\pi(c_{L1}), \pi(c_{L2}))$:
\begin{restatable}[Data independence of cache hierarchies]{cor}{dataindependencehierarchies}\label{cor:dataindependencehierarchies}
Let $c \in \hierarchycachestate$, $b \in \blocks$, and $\pi \in \PiindexTwo$. Then:
\[\pi(\updatehierarchicalcache(c,b)) = \updatehierarchicalcache(\pi(c), \pi(b)).\]
\end{restatable}
\begin{proof}
\[
\begin{array}{rcl}
	&& \pi(\updatehierarchicalcache(c,b))\\
& \overset{(\ref{eq:defupdatecachehierarchy})}{=} & \pi\left(\left(\updatecache_{L1}(c_{L1},b), \begin{cases}
											c_{L2}	&: \classifycache(c_{L1},b)=\textit{true}\\
											\updatecache_{L2}(c_{L2},b)\hspace{-2.5mm}	& : \classifycache(c_{L1},b)=\textit{false}
									           \end{cases}\right)\right)\\
  & = & 	\left(\pi\left(\updatecache_{L1}(c_{L1},b)\right), \begin{cases}
											\pi\left(c_{L2}\right)	&: \classifycache(c_{L1},b)=\textit{true}\\
											\pi\left(\updatecache_{L2}(c_{L2},b)\right)\hspace{-2.5mm}	& : \classifycache(c_{L1},b)=\textit{false}
									           \end{cases}\right)\\								    
 &\overset{(\ref{eq:dataindependenceclassification})}{=} &	\left(\pi\left(\updatecache_{L1}(c_{L1},b)\right), \begin{cases}
											\pi\left(c_{L2}\right)	&: \classifycache(\pi(c_{L1}),\pi(b))=\textit{true}\\
											\pi\left(\updatecache_{L2}(c_{L2},b)\right)\hspace{-2.5mm}	& : \classifycache(\pi(c_{L1}),\pi(b))=\textit{false}
									           \end{cases}\right)\\			
&\overset{(\ref{eq:dataindependence})}{=} &\left(\updatecache_{L1}(\pi(c_{L1}),\pi(b)), \begin{cases}
											\pi(c_{L2})	&: \classifycache(\pi(c_{L1}),\pi(b))=\textit{true}\\
											\updatecache_{L2}(\pi(c_{L2}),\pi(b))\hspace{-2.5mm}	& : \classifycache(\pi(c_{L1}),\pi(b))=\textit{false}
									           \end{cases}\right)\\										  					           
 &\overset{(\ref{eq:defupdatecachehierarchy})}{=}  & \updatehierarchicalcache((\pi(c_{L1}),\pi(c_{L2})), \pi(b))\\
 & = & \updatehierarchicalcache(\pi(c), \pi(b))
\end{array}
\]

\end{proof}

A similar statement can be made for a generalization of the classification function to hierarchical caches that distinguishes L1 from L2 cache hits.
We also note that the above statements could be generalized to multi-level cache hierarchies.

\subsection{Constructive Definition of $\symupdatecache$}

Here, we provide a constructive definition of $\symupdatecache$ that satisfies:
\begin{equation*}
	\conc(\symupdatecache((\symc, \vi), \symb)) = \updatecache(\conc(\symc, \vi), \sem{\symb}(\vi))
\end{equation*}

We define $\symupdatecache$ similarly to its concrete counterpart based on a symbolic update of set states:
\begin{multline*}
	\symupdatecache((\symc, \vi), \symb)) := \\
	\symc[\cacheindex(\sem{\symb}(\vi)) \mapsto \symupdateset(\symc(\cacheindex(\sem{\symb}(\vi))), \vi), \symb)]
\end{multline*} 

We define $\symupdateset((\syms, \vi), \symb)$ employing the concrete update:
\begin{equation*}
\symupdateset((\syms, \vi), \symb) :=
\left(\lambda l\in \lines. \left\{\begin{aligned}
	\symb	& : \textit{if } c'(l) = \sem{\symb}(\vi)\\
	\symb'	& : \textit{if } (c'(l), \symb') \in \mathit{BlockPairs}_{c, \syms}\\
	\epsilon	& : \textit{otherwise}
	\end{aligned}\right\}, c'.ps\right)
\end{equation*}

where $c = \conc_\set(\syms, \vi)$ and $c' = \updateset(c, \sem{\symb}(\vi))$ and
\[
\mathit{BlockPairs}_{c, \syms} = 
	\{ 
		(c(l), \syms(l)) 
		\mid  
		l \in \lines, 
		c(l) \neq \varepsilon		
	\}.
\]

\section{Further Experimental Results}

\subsubsection*{Non-warping Simulation versus Traditional Trace-based Cache Simulation}
We evaluated warping performance using our own non-warping simulation as a baseline.
To confirm that our baseline is reasonable, we also compare non-warping simulation to the traditional trace-based cache simulator Dinero~IV~\cite{Edler1999}.
We simulate the L1 cache of the test system with Dinero~IV, but with LRU replacement since Dinero~IV does not support Pseudo-LRU. 
Note that the Dinero~IV simulation times include the trace generation with QEMU~\cite{Bellard2005}.

Figure~\ref{fig:dinero-vs-nonwarping} shows the speed-up of non-warping simulation compared to Dinero~IV.
Although Dinero~IV employs many optimizations, non-warping simulation is faster than Dinero~IV for most kernels.
We believe that the main reason behind this is the overhead related to the retrieval of memory access traces, which is more efficient in the implementation of non-warping cache simulation based on our tree representation of SCoPs.

Considering the speedup of warping simulation compared to non-warping simulation, we conclude that warping provides significant performance benefits over Dinero~IV when we are simulating the L1 cache of the test system. %

\begin{figure}
    \centering
	\includegraphics[width=0.7\linewidth]
	{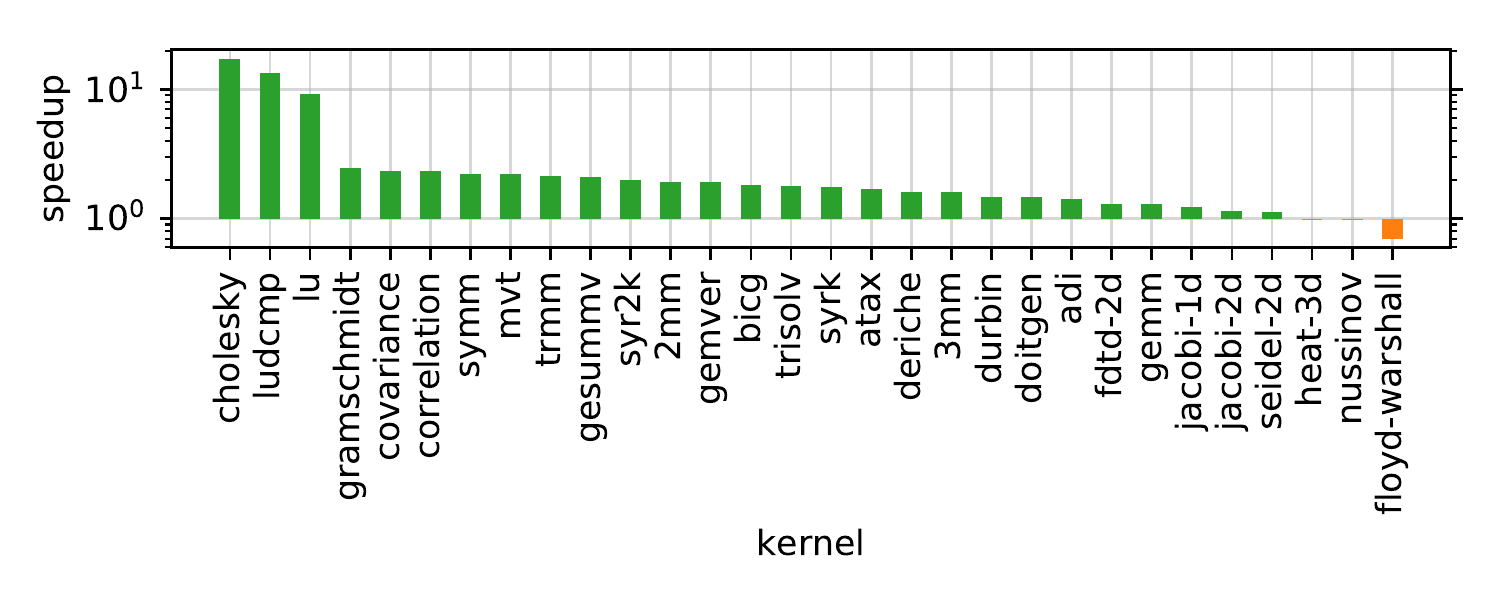}
    \caption{Speedup of L1 non-warping simulation compared to Dinero~IV for problem size L.}
    \label{fig:dinero-vs-nonwarping}
\end{figure}

\subsubsection*{Comparison with Measurements on Actual Hardware}
In Figure~\ref{fig:papi-accuracy-l1} we compared the accuracy of Dinero~IV, warping simulation, and HayStack w.r.t. the measured number of misses on the actual hardware.
Those experiments were performed for the large problem size.
We performed the same experiments also for the small and medium problem sizes of PolyBench.
Figures~\ref{fig:papi-accuracy-l1-small} and~\ref{fig:papi-accuracy-l1-medium} show the corresponding results.
  
Somewhat surprisingly, for these smaller problem sizes, more marked differences are visible between HayStack and the other two approaches.  
We speculate that for the large problem size, a larger share of the memory accesses can easily be classified as misses, because the workload is very clearly too large for the L1 cache.
For the small and particularly the medium problem size, a larger share of accesses might be ``at the edge'', requiring an accurate cache model for correct classification.
However, this needs further investigation.
 
\newpage
\begin{figure}
    \centering
	\includegraphics[width=0.955\linewidth]
	{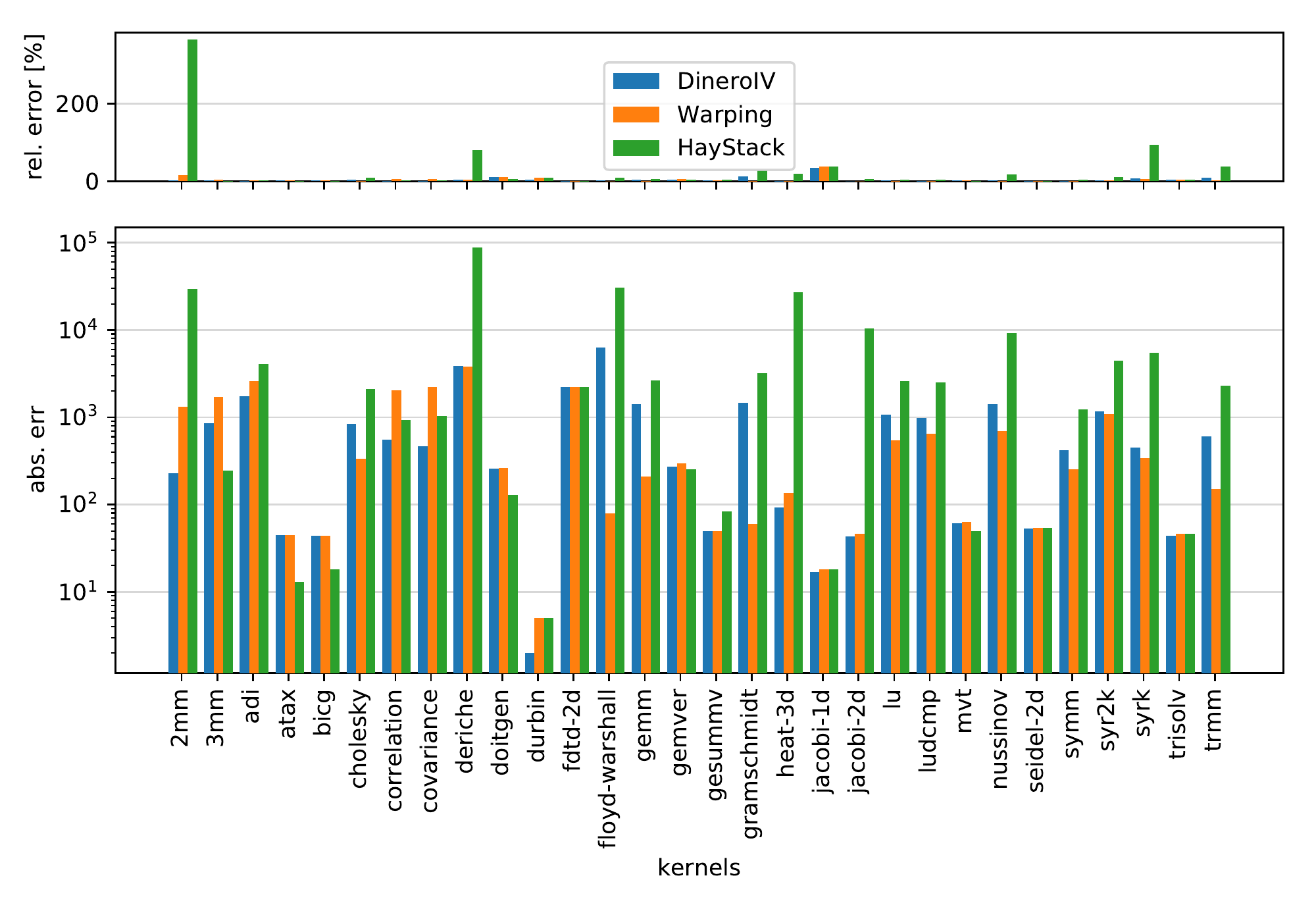}
   \caption{Accuracy relative to measurements on the actual hardware using PAPI for the small problem size.}
   \label{fig:papi-accuracy-l1-small}
\end{figure}

\begin{figure}
    \centering
	\includegraphics[width=0.955\linewidth]
	{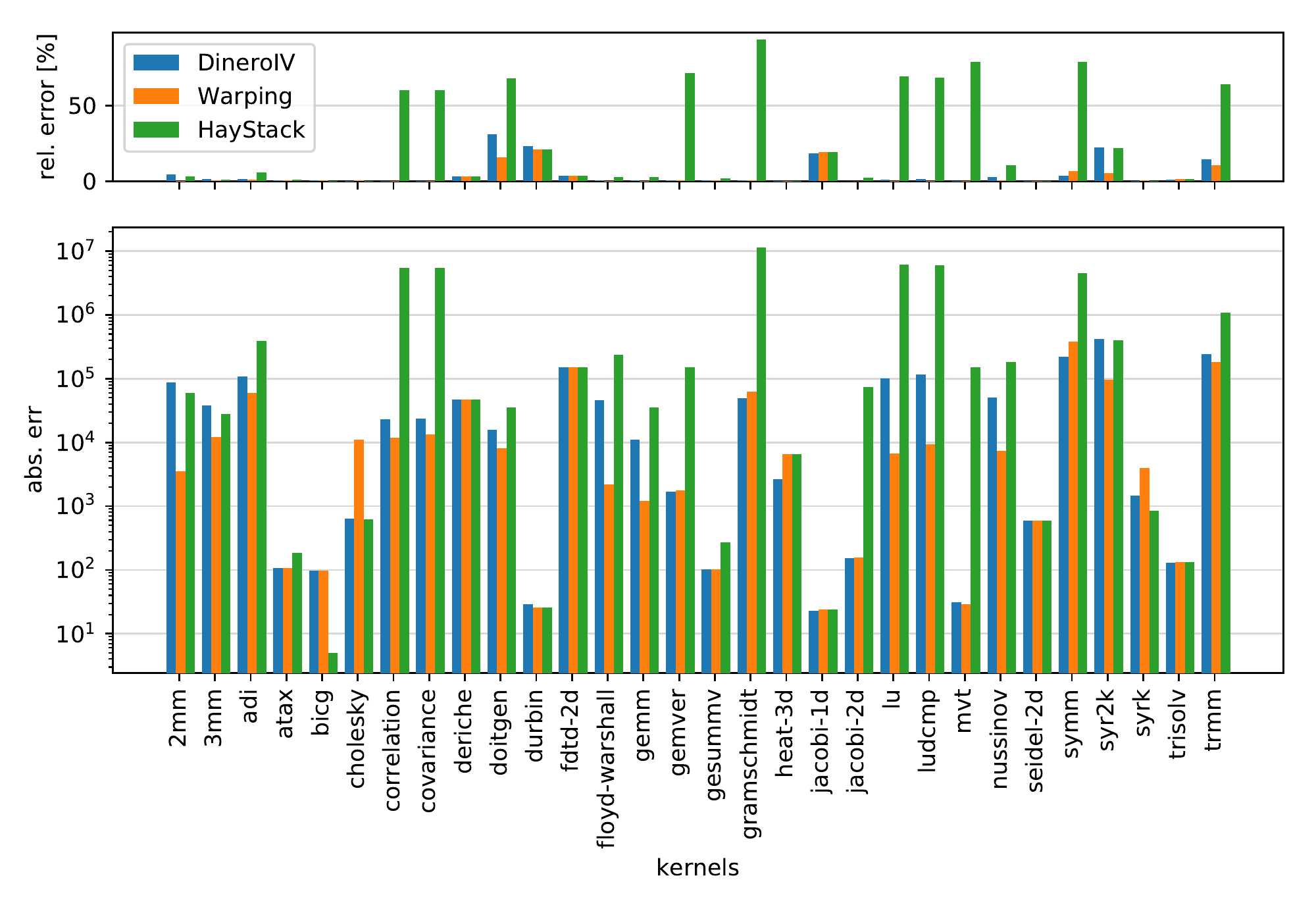}
    \caption{Accuracy relative to measurements on the actual hardware using PAPI for the medium problem size.}
    \label{fig:papi-accuracy-l1-medium}
\end{figure}

\fi

\end{document}